\documentclass{elsarticle}

\usepackage{lineno,hyperref}
\modulolinenumbers[5]
\usepackage{hyperref}       
\usepackage{url}            
\usepackage{amsfonts}       
\usepackage{nicefrac}       
\usepackage{amsmath} 
\usepackage{graphicx}
\usepackage{xcolor}

\graphicspath{ {./figures/} }

\begin{document}

\begin{frontmatter}

\title{A differential Hebbian framework for biologically-plausible motor control}

\author[]{Sergio Verduzco-Flores\corref{corresponding}}
\cortext[corresponding]{Corresponding Author.}
\ead{sergio.verduzco@gmail.com}

\author{William Dorrell}
\author{Erik De Schutter}

\address{Computational Neuroscience Unit,
  Okinawa Institute of Science and Technology,
  Okinawa, Japan.}


\begin{abstract}
In this paper we explore a neural control architecture
that is both biologically plausible, and capable of fully autonomous learning.
It consists of feedback controllers that learn to achieve a
desired state by selecting the errors that should drive them. This selection
happens through a family of differential Hebbian learning rules that, through
interaction with the environment, can learn to control systems where the error
responds monotonically to the control signal. We next show that in a more
general case, neural reinforcement learning can be coupled with a feedback
controller to reduce errors that arise non-monotonically from the control
signal. The use of feedback control can reduce the complexity of the reinforcement
learning problem, because only a desired value must be learned, with the
controller handling the details of how it is reached. This makes the function to
be learned simpler, potentially allowing learning of more complex actions. 
We use simple examples to illustrate our approach, and discuss how it could be 
extended to hierarchical architectures.
\end{abstract}

\begin{keyword}
synaptic plasticity\sep motor control\sep reinforcement learning\sep 
feedback control
\end{keyword}

\end{frontmatter}


\section{Introduction}

Understanding animal motor control holds the promise of improving therapies for
people with motor deficits. Moreover, complex motor control in animals remains
superior to current artificial systems, so insights from animal motor control may 
one day improve state-of-the-art artificial control.
To reach such understanding, we need models that obey strong biological
plausibility constraints, but still perform increasingly complex
motor tasks.

We believe that serious attempts at biological plausibility should consider the
following points:

\begin{itemize}
\item Modeling the full sensorimotor loop with a controller that only uses
neurons. Learning consists of adjusting the weights of their synaptic connections.
\item Learning rules use only information locally available at the postsynaptic 
neuron.
\item The agent learns as its body interacts in real time with the environment.
Rather than relying on labeled data, learning takes advantage of correlation 
between signals, and reinforcement learning mechanisms.
\item Transmission delays and response latencies should be considered.
\item No element of the model goes against current consensus in neuroscience.
\end{itemize}

We are not aware of motor control models that follow all these guidelines,
and only a few follow most of them. This is because complications arise in
biological models. The worst complication may be the one 
recently dubbed as the {\it supraspinal pattern formation problem}
\cite{bizzi_motor_2020}: how are the
spinal cord components coordinated in time to generate goal-directed movements?
A closely related complication is that many motor patterns may achieve the same
motor outcome. This was originally known as the {\it DOF problem}
\cite{bernstein_co-ordination_1967}, or more commonly as the {\it redundancy problem}.

In this paper we lay a framework for motor control
that incorporates all the biological constraints above, while offering a
viable solution to supraspinal pattern formation and redundancy. The key is to
cast the problem in terms of finding the input-output structure of a
Multiple-Input Multiple-Output (MIMO) feedback control system 
\cite[Ch.18]{seborg_process_2016}, or in other terms, solving the input-output
decoupling problem \cite{nijmeijer_input-output_1990}. This problem is about
choosing the right actuator (controller output) in order to reduce the error for
each controlled variable (controller input). Its main complication is that
the actuators may affect several controlled variables, so using one of them
to control a variable may cause unwanted interference in 
the state of other variables. In engineering systems this is usually addressed
during the design stage, but at least in primates this is likely learned
through experience. 

The approach we use to find the input-output structure in MIMO feedback control
relies on learning {\it sensitivity derivatives} using differential Hebbian
learning with synaptic competition. The sensitivity derivatives are the values
$de_i/dc_j$, where $\mathbf{c}=[c_1, \dots, c_N]$ is the output vector produced
by the controller in order to reduce an error vector 
$\mathbf{e}=[e_1, \dots, e_M]$.

We will find that this can be an effective
solution, but that it fails in cases where the relation between input and output
changes for different contexts. To handle this scenario we will combine our
feedback controller with a variant of the actor-critic architecture, which will
allow it to self-configure for handling different contexts.


Most animal motor control models use a fixed input-output
structure (e.g.
\cite{hayashibe_synergetic_2014,kawato_computational_1992,porrill_adaptive_2013,todorov_direct_2000}).
When asking how are motor errors defined and used, they
assume that this is either genetically determined, or adjusted through an
internal model. 
There is extensive evidence for the presence of internal forward models 
predicting the consequences of motor commands, and that they
adapt when those consequences change due to perturbations 
(e.g. \cite{miallForwardModelsPhysiological1996, mcnameeInternalModelsBiological2019,
tanakaCerebroCerebellumLocusForward2020}). 
It is thus often assumed that motor corrections arising from errors
are caused by a correction to a forward model 
\cite{jordan_forward_1992, wolpert_internal_1995}. An alternative that is not often
considered is that the motor corrections are independent from the corrections to
the forward models. Recent experiments suggest that this may be the case:
errors in the sensory domain seem to generate motor
corrections without using forward models 
\cite{hadjiosif_did_2021}. 

Sensitivity derivatives constitute a linear forward model, not of the
system being controlled, but of the errors, which contain information about the
desired outcome. As will be shown later, estimating a form of these values will
directly produce error corrections, and adjust the control structure of the system.
In contrast, approaches using internal models of the system being controlled
(called the {\it plant}) need to train such
models, and make them produce corrections; this usually requires a
pre-existing control structure (e.g.  
\cite{porrill_recurrent_2004,miyamoto_feedback-error-learning_1988}), or a form
of error backpropagation \cite{jordan_forward_1992}. 

In addition of not depending on a forward model, the model we present consists
entirely of neurons.
Four control architectures using biologically-plausible neural networks are 
well known \cite{rokni_neural_2009}, each presenting its own 
strengths and limitations. Direct inverse learning 
\cite{kuperstein_neural_1988} uses the correlations between muscle outputs and
afferent inputs in order to approximate an inverse function that maps from
desired afferent inputs to the muscle activity that produces them. A major
drawback is that the relation between muscle activity and afferent inputs
may not be invertible (e.g. many muscle activities producing the same results).

Distal supervised learning \cite{jordan_forward_1992} is another neural network
architecture for control. It relies on both forward and inverse models of the
plant. In order to produce learning signals for the inverse model, the errors in
the forward model must be backpropagated.
Feedback error learning \cite{miyamoto_feedback-error-learning_1988} also uses 
an inverse model of the plant, but instead of relying on a forward model, it
uses the error of a closed-loop feedback controller to train it. This avoids the
need of a forward model as in distal supervised learning, but it relies on
a pre-existing closed-loop controller.

The fourth architecture is Reinforcement Learning (RL), which avoids the limitations
of the other architectures, but is generally slower to find a solution. Given the
close ties between RL and differential Hebbian learning 
\cite{kolodziejski_asymptotic_2008}, it is
interesting to ask whether the correlations between inputs and outputs to the
controller can be used to obtain a control law that is adaptive and biologically
plausible. As far as we know this has not been attempted in order to obtain the
sensitivity derivatives in closed-loop control (cf. 
\cite{kolodziejski_mathematical_2008}). 

We are aware of one single work concerned with finding sensitivity derivatives
in a biologically plausible manner.
In \cite{abdelghani_sensitivity_2008} the sensitivity derivatives are
represented as the firing rates in a separate network doing expansive recoding
of appropriate context variables, together with a variant of the LMS learning
rule. The authors in this work were unable to represent the sensitivity
derivatives without using fast weight transport (which is biologically
implausible), so they had to represent them as firing rates.
The approach that we will present below is capable of using
synaptic weights to represent something analogous to the 
sensitivity derivatives. This permits memory of the learned variables.
Moreover, we show that in a feedback architecture many learning rules can
achieve this, with approaches within and outside of the RL framework.

There are in fact four models presented in this paper. 
In the Methods we first show a heuristic derivation of the differential Hebbian
learning rules, and then describe each of the four models.

The learning rules we derive allow a proportional feedback control
system to adjust so as to reduce an arbitrary error, as long as the error and
the motor commands have a monotonic relation. In other words, the motor command
should not cause the error to increase in one context, and to decrease in a
different one. 

The first model we present is a direct application of these learning rules to
find the input-output structure of high-dimensional linear plants with varying
levels of redundancy in the actuators. From this we will observe that the
tolerance to redundancy is on par with some offline analytical approaches.

The second model uses one of our learning rules to control the angle of a pendulum.
The pendulum is a 2-dimensional plant, so finding the input-output structure of
a controller is not particularly hard. On the other hand, even if our feedback
controller has the right input-output structure, it only provides proportional
control, which is insufficient to deal with the pendulum's momentum. We thus
modify the architecture of the feedback controller to incorporate velocity in
the error through the {\it input correlation} learning rule
\cite{porr_strongly_2006}, resulting in a biologically-plausible, self-configuring
proportional-derivative controller. 

The third model illustrates a way that the limitation of monotonic errors
mentioned above may be overcome. We again control a pendulum, but the signal
that represents its angle has a discontinuity as the pendulum completes a full
revolution, something that negative feedback control cannot compensate by
changing its input-output structure. We thus enhance the controller with a
{\it critic} component that indicates which angle representation to use for
each context.

The RL methods we use in the third model are fairly standard: a neural
implementation of TD-learning \cite{schultz_neural_1997}, and reward-modulated
Hebbian learning. However, the times at which the reward-modulated Hebbian rule
updates are non-standard. The fourth model in this paper is meant to show that
this is not arbitrary, as it can be useful in solving temporal credit assignment
problems. To this end, in the fourth model a very simple controller uses
reward-modulated Hebbian learning to solve the inverted pendulum problem.

The Results section illustrates the performance of the four models described in
the Methods.

All models in this paper are meant to illustrate and provide proof-of-concept
for the ideas in our approach to motor control. Application to the control of a
more realistic biological system is presented in a subsequent paper
\cite{verduzco-flores_adaptive_2021}.



\section{Methods}

Simulations for all models were implemented
in the Draculab neural simulator \cite{verduzco-flores_draculab:_2019}. 
The values for parameters appearing in this paper are
reported in \ref{app:parameters}.
The Supplementary Material to this paper includes the source
code, where these and other parameter values are contained within Python
dictionaries.

\subsection{Differential Hebbian learning rules}
\label{sub:rules}
\begin{figure}
  \centering
  \includegraphics[width=0.5\linewidth]{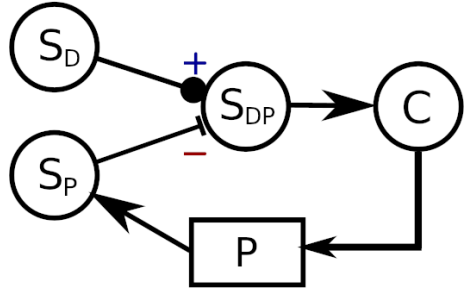}
  \caption{ A negative feedback controller. The circles
          represent populations of neural units 
          whose output is a scalar value between 0 and 1 (e.g.  firing rate 
          neurons). Excitatory connections end with a closed circle, inhibitory
          connections with a bar. Connections with arrows can have inhibitory
          and excitatory components.}
\label{fig:neg_feedback}
\end{figure}

Consider a negative feedback controller as depicted in figure
\ref{fig:neg_feedback}. The goal of this controller is to make the activity of
the $S_P$ neural population equal to that of a population $S_D$ that provides
desired values. The output of the $S_{DP}$ population is an $M$-dimensional
error vector $\mathbf{e} = \left[e_1, \dots, e_M \right]$. Population $C$
contains $N$ units whose activity is in the vector
$\mathbf{c} = \left[c_1, \dots, c_N \right]$. 
We assume that 
\begin{equation} \label{eq:unit_dynamics}
\tau_c \dot{c}_i = \sigma \left(\sum_{j=1}^M \omega_{ij} e_j \right) - c_i,
\end{equation}
where:
\begin{equation} \label{eq:sig}
\sigma(x) = \frac{1}{1 + \text{e}^{-\beta(x - \eta) }}.
\end{equation}
The parameter $\tau_c$ is a time constant controlling the response latency of
the controller's units. $\omega_{ij}$ is the synaptic weight for the connection
from $e_j$ to $c_i$.
$\beta$ is the ``slope'' of the sigmoidal activation function, 
and $\eta$ is its ``threshold''.

For this derivation we assume that internal connections within neurons of the 
same population have a negligible effect (although this restriction is not 
necessary \cite{verduzco-flores_adaptive_2021}).
All synaptic connections are static,
except those from $S_{DP}$ to $C$, where we assume all-to-all connectivity.
The result of this subsection will be two different alternatives for learning
in the weights $\omega_{ij}$ of these connections. These learning rules are in the
following two equations:

\begin{equation} \label{eq:rga}
\dot{\omega}_{ij}(t) = -\alpha \Big(\dot{e}_j(t) - \langle \dot{e}(t)
\rangle \Big)
\Big( \dot{c}_i(t-\Delta t) - \langle \dot{c}(t-\Delta t) \rangle \Big),
\end{equation}

\begin{equation} \label{eq:mixed}
\dot{\omega}_{ij}(t) = -\alpha \Big(\ddot{e}_j(t) - \langle \ddot{e}(t) 
\rangle \Big) \Big(\dot{c}_i(t-\Delta t) - 
\langle \dot{c}(t-\Delta t) \rangle \Big).
\end{equation}

In both equations $\alpha$ is a learning rate parameter, and $\Delta t$ is a
parameter that approximates the time required for a control signal to propagate
around the loop. In other words, a change $\dot{c_i}$ in one of the
controller outputs will roughly take $\Delta t$ seconds to manifest as a change
$\dot{e}_j$ or $\ddot{e}_j$ in the errors. The brackets used in the equations
indicate an average over all the units in the same population:
$\langle \dot{e}(t) \rangle \equiv \frac{1}{M} \sum_k \dot{e}_k(t)$,
$\langle \dot{c}(t) \rangle \equiv \frac{1}{N} \sum_k \dot{c}_k(t)$.

Rather than coming from a loss function, the rules in equations \ref{eq:rga} and
\ref{eq:mixed} are the result of an informal heuristic procedure, which is described
next. 

First, we should notice that setting the $\omega_{ij}$ weights so they minimize the
error is in fact solving the input-output structure problem for the
proportional controller in figure \ref{fig:neg_feedback}. To reduce the error,
we want $e_j$ to activate $c_i$ when $c_i$'s activity reduces
$e_j$. This is tantamount to having the weight $\omega_{ij}$ from
$e_j$ to $c_i$ be proportional to the negative of their sensitivity derivative:
\begin{equation}
\omega_{ij} \propto - \partial e_j / \partial c_i. 
\end{equation}
In this way the errors that arise will trigger an action to cancel them.

We remain agnostic about the properties of the plant and how its state is
transformed into perceived values in $S_P$, but we assume that the sensitivity
derivatives maintain their signs, and that the propagation constant 
$\Delta t$ does not change significantly.

Our aim is not to
have accurate estimates $\omega_{ij} \approx - \partial e_j / \partial c_i$, but
rather to give $\omega_{ij}$  a magnitude that is
appropriate for feedback control. The Relative Gain Array (RGA) criterion 
\cite{bristol_new_1966} is a classical method to achieve this, inspiring some
of the procedure below (see \ref{app:rga} for more details), but due to
reasons of biological plausibility we do not exactly implement it.

The most straightforward way to obtain estimates for $\omega_{ij}$ may be to let
the system settle into a fixed point, and then
to produce a perturbation $\Delta c_i$, resulting in a change $\Delta e_j$ for
the errors. Weights can be adapted as $\Delta \omega_{ij} \propto -\Delta e_j /
\Delta c_i$. While this is feasible, and suggestive of possible learning taking
place in unborn mammals (e.g. \cite{brumleyDevelopmentalPlasticityCoordinated2015,
hamburgerAnatomicalPhysiologicalBasis1973})
we are interested in the case of online learning, where
$\omega_{ij}$ is adapted during performance of a behavior.

A simple approach to online learning 
is to use the correlation of the
first derivatives. This provides a measure of whether $c_i$ and $e_j$
change together, in a way that is invariant to their mean values.
The resulting learning rule is: 
\begin{equation*}
\dot{\omega}_{ij}(t) = -\alpha \dot{e}_j(t) \dot{c}_i(t-\Delta t).
\end{equation*}
where $\Delta t$ is an approximation to the time it takes $c_i$ to change the
perceived error $e_j$, and $\alpha$ is a learning rate.

This approach has three main limitations. Firstly, during behavior the
whole $\dot{\mathbf{c}}$ vector acts as the perturbation, so it is unclear
which of the $c_i$ units is responsible for an observed change $\dot{e}_j$.
Secondly, an observed change $\dot{e}_j$ may not be the effect of any recent
$\dot{c}_i$ change, but rather part of the normal flow in state space for the
current state. Thirdly, the magnitudes
$\frac{\partial e_j}{\partial c_i}$ are functions of 
the state $x_P$ of the plant (and potentially of $\mathbf{c}$), so they could 
change sign for different contexts.

We will address each of these 3 limitations. In short, to mitigate the first
one we will introduce synaptic competition in the learning rule, and the
second one will be handled by introducing a second order derivative, turning
equation \ref{eq:rga} into equation \ref{eq:mixed}.
The third limitation is more subtle, and will require that we divide our approach
into the case when $\frac{\partial e_j}{\partial c_i}$ does 
not change
sign (monotonic control), and the case when the sign changes. Nonmonotonic
control will be handled by introducing a reinforcement learning mechanism
that changes the configuration of the controller in different regions
of state space.

Next we introduce synaptic competition in the learning rule.  
Using the term 
$( \dot{c}_i - \langle \dot{c} \rangle )$
rather than $\dot{c}_i$ we expect that on average, weights corresponding 
to the largest sensitivity derivatives will be enlarged, whereas weights
with below-average sensitivity derivatives will shrink. This should allow
for errors to be reduced by the $c_i$ units that have the largest effect on
them. Notice that lateral connections among the $C$ units
is what make the $c_k$ values locally available.

As explained in \ref{app:rga}, the RGA criterion relies on a vector
perturbation $\Delta \mathbf{c}^j$ that alters only one of the errors (e.g. 
$\Delta e_l = 0$ for $l \neq j$). The gain of this perturbation is used
to select the inputs to the controller, with the idea that when the
$e_j$ error arises, the controller response that causes the least
interference should be aligned with $\Delta \mathbf{c}^j$. A simple, 
biologically plausible version of this approach does not seem
likely, but a further application of synaptic competition may achieve a 
similar purpose.

By using $(\dot{e}_j - \langle \dot{e} \rangle)$ in the learning equation
rather than just $\dot{e}_j$ we may select only the controller units that have
a large effect on $e_j$. Together with the previous use of synaptic competition,
this creates a sparser response that hopefully mitigates the creation
of new errors when reducing $e_j$. Introducing this change leads us to equation
\ref{eq:rga}.

The rule in equation \ref{eq:rga} can effectively configure the feedback loop of
simple MIMO systems (see section \ref{sub:linear_control}), but 
it can further be improved. In particular, we may replace $\dot{e}_j$
by $\ddot{e}_j$ in order to remove the effect of changes where $\dot{e}_j$
comes from momentum in the plant rather than the action of a controller.
The resulting rule is also what we would obtain from
the previous discussion, if we had assumed that a change $\dot{c}_i$ in the
output produced a response $\ddot{e}_j$ in the $j$-th error.
This simple change leads to equation \ref{eq:mixed}.


Equation \ref{eq:mixed} is better suited for the control of systems where the
plant's dynamics are important. For example, 
$\mathbf{c}$ may be a force, and $\mathbf{e}$ a displacement or a velocity,
so if the plant follows Newton's laws we should expect
the correlations to appear among derivatives of different orders.

For the models in this paper, equations \ref{eq:rga} and \ref{eq:mixed}
include two additional modifications: connection weights do not change sign, and
the sum of weights remains constant. In order to maintain the initial sign of
the weights, the whole learning equation is multiplied by $\omega_{ij}$, a
strategy called ``soft weight-bounding''. To maintain the sum constant,
a normalization term was included in the equation.

The normalization term leveraged two requirements.
First, that all weights from projections
starting from the same $S_{DP}$ unit should add to $w_{sa}$. Second, the
sum of all $S_{DP}$-to-$C$ weights terminating in the same $C$ unit should add
to $w_{sb}$. Let $\zeta_{j}^{sa} \equiv w_{sa} / \sum_k \omega_{kj}$, and
$\zeta_{i}^{sb} \equiv w_{sb} / \sum_k \omega_{ik}$. Equations \ref{eq:rga} and
\ref{eq:mixed}, using soft-weight bounding and normalization, had the form:

\begin{equation} \label{eq:w_norm}
\dot{\omega}_{ij} = \omega_{ij} \left( \Omega +
\alpha \lambda \left[ 1 - \frac{\zeta^{sa}_{j} + 
\zeta^{sb}_{i}}{2} \right] \right),
\end{equation}
where $\Omega$ is the right-hand side of either equation \ref{eq:rga} or
equation \ref{eq:mixed}, and $\lambda$ is a scalar parameter.  This type of 
normalization is meant to reflect the competition
for resources among synapses, both at the presynaptic and postsynaptic level.

To obtain the derivatives used in the learning rules in a biologically-plausible
manner, we approximated rates of change as the difference of two first-order
low-pass filters. We assumed $\dot{c}(t) \propto \mathbf{c}_{fast} - 
\mathbf{c}_{slow}$, where 
\begin{align} \label{eq:fast_low_pass}
\tau_{f}\dot{\mathbf{c}}_{fast} = c - \mathbf{c}_{fast}, \\  
\tau_{s}\dot{\mathbf{c}}_{slow} = c - \mathbf{c}_{slow},
\label{eq:slow_low_pass}
\end{align}
and $\tau_f \ll \tau_s$. 

Elements like $\mathbf{c}_{fast}$ and $\mathbf{c}_{slow}$ can come from feedback
connections (cf. Eq. 19 in \cite{lim_balanced_2014}), but it is also possible
that they could represent the concentration of molecules involved in the
postsynaptic depolarization, and the subsequent chemical cascades. For example,
intracellular calcium concentration has been described as a possible indicator
of firing rate, using leaky integrator dynamics \cite{helmchen_dendrites_1999}.

Equations \ref{eq:rga}, and \ref{eq:mixed} are by no means the only options to
self-configure a feedback loop. In \ref{app:alternative} we present
two alternative derivations. The first one is meant to explore whether
the established reinforcement learning methods are adequate for this
problem. The other derivation in \ref{app:alternative} is based on stability
considerations. It is shown that neither of those rules was more 
effective than equations \ref{eq:rga} and \ref{eq:mixed}.

\subsection{Linear MIMO system controller}
\label{sub:linear}

The first application of our learning rules (Eqs. \ref{eq:rga}, \ref{eq:mixed},
\ref{eq:w_norm}) is in the control of a linear plant. 

\subsubsection{The controller}
Unit activities are non-negative, but the controller needs to know the sign of
the error. Two basic options for this are:
1) to have units in the $S_{DP}$ population signal negative
values as deviations below a baseline level, and positive values as deviations
above this level; or 2) to have two separate populations, one for each sign of
the error. In other words, this last option amounts to have one population with
activity monotonically related to
$\max(\mathbf{0}, \mathbf{s}_D - \mathbf{s}_P)$, and another population whose
activity is a monotonic function of 
$\max(\mathbf{0}, \mathbf{s}_P - \mathbf{s}_D)$, where 
$\mathbf{s}_D,\mathbf{s}_P$
are the activities of $S_D$ and $S_P$, respectively.

\begin{figure}
  \centering
  \includegraphics[width=0.5\linewidth]{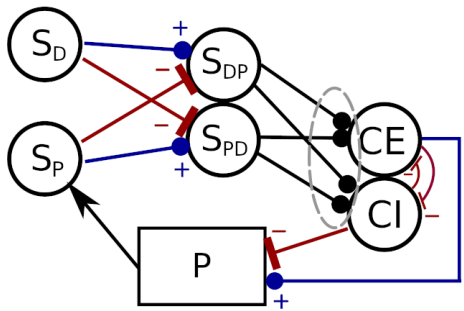}
  \caption{ Negative feedback controller with dual populations, and synaptic 
        weights that are either excitatory or inhibitory. 
        Connections inside the gray
        dashed oval are adjusted using the learning rules of section
        \ref{sub:rules}. Blue circles indicate excitatory connections,
        red bars inhibitory connections, and arrows are afferent inputs that 
        can be excitatory or inhibitory, but do not change sign. 
        }
\label{fig:bio_feedback}
\end{figure}

We believe our learning rules can work with either solution, 
but for the purpose of this paper we found the second option to
be more appropriate. Accordingly,
we modified the architecture of figure \ref{fig:neg_feedback}
by separating $S_{DP}$ and $C$ into two separate populations each, resulting
in the architecture of figure \ref{fig:bio_feedback}.
In this figure $S_{DP}$ is excited
by $S_D$, and inhibited by $S_P$. We assume this inhibition happens through
local interneurons, not explicitly modeled. $S_{PD}$ receives the opposite
activation of $S_{DP}$, so that when an error has a positive sign (e.g
$s_D > s_P$), a unit in $S_{DP}$ will activate, whereas a negative error will
activate a corresponding unit in $S_{PD}$. In this way the error activities
$e_j$ will always be positive, but also capable of signaling errors in
either direction. Having two separate populations to represent sensory events,
one being inhibited while the other is excited, is termed {\it dual
representation} in this paper. 

Units in the $S_P, S_{DP}$, and $S_{PD}$ populations
use sigmoidal units whose activity follows dynamics like those in equation
\ref{eq:unit_dynamics}. To increase biological
plausibility and help avoid synchronization, the threshold and slope of
sigmoidals in these 3 populations used heterogeneous values, with a random
component that ranged from -10\% to 10\% of their original value.

It can be shown that using linear units and a learning rule as in equation
\ref{eq:rga} in a feedback controller allows convergence to fixed points with
non-zero error (see \ref{app:nonconvergence}). To avoid
this the architecture of figure \ref{fig:bio_feedback} uses $CE$ and $CI$ units
that output the integral of their inputs, in addition to displaying intrinsic
noise. Their equations are:
\begin{align} \label{eq:c_2d}
& \tau_x \dot{x}(t) = x(t)(I_{DP} + I_C x(t))(1 - x(t)), \\
& \tau_c \dot{c}(t) = x(t) - c(t) + \zeta. \label{eq:c_2d_b}
\end{align}
$I_{DP} \equiv \sum_k \omega^{PD}_k s_k$, representing the sum of inputs from 
$S_{DP}, S_{PD}$ times their synaptic weights. 
$I_C \equiv \sum_k \omega^{C}_k c_k$ is the sum of inputs arising 
from $CE, CI$ times their weights;
$\tau_x, \tau_u$ are time constants, and $\zeta$ is a white noise process.

Integration of inputs is a basic neuronal computation
\cite{izhikevich_neural_2000}. In equation \ref{eq:c_2d} this integration is
combined with soft weight bounding to keep the integration factor $x$ between 0
and 1. The term $(I_{DP} + xI_C)$ is an input sum where ``lateral'' inputs 
are reduced for small $x$ values. This avoids ``winners-take-all'' dynamics
in $C$. Equation \ref{eq:c_2d_b} simply slows down convergence of the firing
rate to the integral, and adds noise.  This Langevin equation
was solved using the Euler-Maruyama method, whereas all the other equations were
solved with the forward Euler method.

One undesired consequence of soft weight-bounding as in equation \ref{eq:c_2d},
is that when $x(t)$ is very close to 0 or 1 the inputs have little effect, and
the unit may stay stuck at that value. To avoid this, if $x(t)$ ever surpassed
0.97 its derivative would become $0.9 - x(t)$. Furthermore, to enhance
numerical stability, the derivative of $c(t)$ was clipped if its absolute value
became larger than 1.

\subsubsection{The plant}
The linear plant $P$ is defined by associating each unit $c^e_j$ in 
$CE$ with a
vector $\mathbf{v}_j$, whereas the corresponding unit $c^i_j$ in $CI$ is
associated with $-\mathbf{v}_j$.
The plant's response was updated as:

\begin{equation} \label{eq:linear_plant}
        \tau_p \dot{\mathbf{p}} = \left[
        \sum_j (c^e_j - c^i_j)\mathbf{v}_j \right] - \mathbf{p},
\end{equation} 
where $c^e_j, c^i_j$ are also used to denote the activity of those units. 

The amount of redundancy in the controller can be adjusted through the number of
units in $CE, CI$, and by the specific values of the $\mathbf{v}_j$ vectors.
This information is contained in the connection matrix from $C$ to $P$, denoted
by $W_{CP}$. Notice that the columns of $W_{CP}$ come from the
$\mathbf{v}_j$ vectors.

We used 4 different $W_{CP}$ matrices for our tests. The first one
tests the performance of the learning rules in a system
with no redundancy. Because of dual representation, $W_{CP}$ was
the following block matrix:
\begin{equation}
W_{CP}^{id} = 
\begin{bmatrix}
I_N & -I_N 
\end{bmatrix},
\end{equation}
where $I_N$ is the $N \times N$ identity matrix, and $N$ is the dimension of the
plant.

The second $W_{CP}$ matrix was built by using the vectors of an $N$-dimensional
Haar basis \cite{strang_wavelet_1993} as the $\mathbf{v}_j$ vectors . 
These vectors form an
orthogonal basis with positive and negative entries. It is defined for linear spaces
where the dimension is a power of 2, so we tested the cases where $N$ is equal
to 2, 4, and 8. Quite importantly, all the vectors of the Haar basis have
several non-zero entries, so the action of any unit $c_j$ will affect several of
the plant variables, but the plant should still be controllable.

Let $H_N$ represent the $N$-dimensional Haar matrix where the columns are
normalized to have unit norm. Our second $W_{CP}$ matrix is the following
$N \times 2N$ block matrix:
\begin{equation}
W_{CP}^{Haar} = 
\begin{bmatrix}
H_N & -H_N 
\end{bmatrix}.
\end{equation}

The third $W_{CP}$ matrix we used is meant to increase the redundancy in
$W_{CP}^{Haar}$. To this end we increased the
number of units in the $CE$ and $CI$ populations, from $N$ to $2N$. 
Let $R_N$ be an $N \times N$ matrix whose columns are random vectors
with unit norm. We used the following connection matrix:
\begin{equation}
W_{CP}^{oc} = 
\begin{bmatrix}
R_N & & H_N & -R_N & -H_N 
\end{bmatrix}.
\end{equation}

The fourth matrix, $W_{CP}^{oc2}$, is used to test a worst-case scenario,
where redundancy is high, and controllability is not ensured. In this case $CE$
and $CI$ each had $3N$ units. The $\mathbf{v}_j$ vectors were random vectors with
unit norm.

All the other static connections used either the identity weight matrix $I_N$
($P$-to-$S_P$, $S_P$-to-$S_{PD}$, $S_D$-to-$S_{DP}$), or its negative $-I_N$
($S_P$-to-$S_{DP}$, $S_D$-to-$S_{PD}$).

\subsubsection{Analytical approaches}
\label{subsub:analytical}
In order to evaluate the performance of our learning rules, we compared it with
two analytical approaches. The first one is based on the Moore-Penrose 
pseudoinverse. Let $W_{SC}$ be the connection matrix from $(S_{PD},S_{DP})$ to 
$(CE, CI)$. If we set $W_{SC} = -W_{CP}^{-1}$, then, ignoring the sigmoidal
nonlinearities, the joint action of the controller and the plant would be akin
to the applying the linear transformation 
$W_{SC} W_{CP} = - W_{CP}^{-1} W_{CP} = - I_N$. Therefore, if $W_{CP}$ is
invertible, the controller may be able to achieve decoupled proportional
control. Since $W_{CP}$ may not be invertible, or square, we set $W_{SC}$ as the
negative of the Moore-Penrose pseudoinverse.

The second approach to obtain $W_{CP}$ is the RGA criterion, as described in
\ref{app:rga}. In this procedure the designer personally assigns a controller 
output for each plant variable that requires control. This is done by searching
entries that are close to 1 in the relative gain array matrix. The values
chosen, however, are to some degree arbitrary. For example, these are the RGA
matrices corresponding to the Haar matrices of dimensions 2 and 4:
\begin{align*}
W_{RGA2} &= 
\begin{bmatrix}
.5 & .5 \\
.5 & .5 
\end{bmatrix}, \\
W_{RGA4} &= 
\begin{bmatrix}
.25 & .25 & .25 & .25\\
.25 & .25 & .25 & .25\\
.5 & .5 & 0 & 0 \\
0 & 0 & .5 & .5 
\end{bmatrix}. 
\end{align*}
In order to create $W_{SC}$ connection matrices from the RGA matrices, for each
error in $S_{PD},S_{DP}$ we assigned one $C$ unit. To choose this unit, for each
column in the RGA matrix (corresponding to one error) we chose the row whose
value was closest to one, and had not been chosen before. If a unit $c_i$ in $CE$
was chosen for error $e_j$ in $S_{DP}$ then the connection from $e_j$ to $c_i$
was 1, and otherwise it was zero. $c_i$ also received a -1 connection from the
dual of of $e_j$ in $S_{PD}$. Moreover, a unit $c_i'$ in $CI$ received the same
connections as $c_i$, but with the signs of the weights reversed.
When there were more
rows than columns, rows not chosen corresponded to units in C that were not 
assigned to control an error, and received inhibition (a -1 connection weight)
from all $S_{PD},S_{DP}$ units. 

The RGA matrices came from this expression:
\begin{equation}
W_{RGA} = W_{CP} \otimes (W_{CP}^{-1})^T,
\end{equation}
where $W_{CP}^{-1}$ is the Moore-Penrose pseudoinverse of $W_{CP}$, and
$\otimes$ denotes the element-by-element product.

\subsection{Monotonic pendulum controller}
\label{sub:monotonic_meth}
The second plant model we tested consisted of a pendulum that cannot
rotate across a certain angle. This means it bounces back when approaching 
$\pm\pi$ radians, so the angles stay in the $(-\pi,\pi)$ range.

The pendulum was modeled after a homogeneous rod of 1 kilogram mass, and 50 
centimeters length. 
Gravity was only included for the simulations in the Appendix.
Angular acceleration is equal to a torque divided by
an inertia moment. The torque had four components: 1) torque generated from the
inputs, 2) viscous friction; 3) and 4) extra torque and viscosity appearing when
the angle approached $\pi$. These last two torques prevented the pendulum from
going across the $\pi$ angle, causing it to bounce, and increasing the friction
while it bounced. Denoting these torques as $\tau_3, \tau_4$, their equations
are:
\begin{align*}
        \tau_3 &= -0.001\tan\left( (\theta\%(2\pi))/2)\right)^3, \\
        \tau_4 &= -\frac{0.05 \ \dot{\theta}}
        {\left[(\theta+\pi)\%(2\pi) + 10^{-5} \right]^2};
\end{align*}
where $\theta$ is the angle, and \% represents the modulo operator.

\begin{figure}
  \centering
  \includegraphics[width=0.5\linewidth]{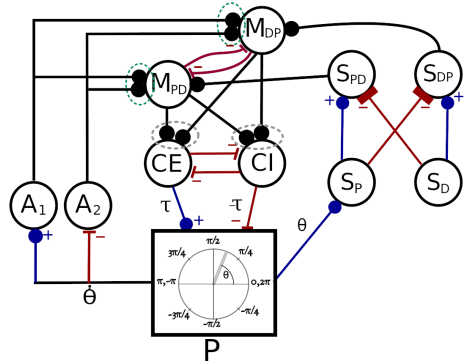}
  \caption{Basic architecture for control
          of a pendulum. Each circle represents a single neuron, whereas the
          square represents the plant $P$. Blue connections are excitatory,
          red ones are inhibitory. $\theta$ represents the current angle in 
          radians, whereas $\dot{\theta}$ is the angular velocity.
          $\dot{\theta}$ is transformed into positive values by logarithmic
          units in the $A$ population (equation \ref{eq:log}). Both units in the $M$
          population receive all $A$ signals. The connections from $A$ to $M$ (green
          dotted ovals) evolve following the input correlation rule, and the connections
          from $M$ to $C$ units (gray dotted ovals) evolve using the rule from equation
          \ref{eq:mixed}. The output of the $C$ units is mapped into either a positive or
          a negative torque ($\tau$). 
          }
\label{fig:arch_pend}
\end{figure}

As explained in section \ref{sub:nonlinear}, 
the architecture of figure \ref{fig:arch_pend} is used for pendulum control.
$C$ uses two units, one providing clockwise, and another counterclockwise torque.
The value in $S_D$ represents a given angle, and the task is to move the pendulum
to that angle so activity in $S_P$ and $S_D$ can be equal. 
$S_D$ does not specify a desired velocity. In order to
adaptively incorporate the velocity information into the control loop we
introduced a population $M$ receiving the 
afferent activity $A$, consisting of the angular velocity $\dot{\theta}$ in its 
non-negative (dual) representation. In addition, each $M$ unit received one
error signal, either $s_{DP}$, or $s_{PD}$. The $M$ units used the
\emph{input correlation} rule \cite{porr_strongly_2006} (equation
\ref{eq:inp_corr}) to potentiate
angular velocity inputs that correlate with their error input.
This allows $M$ to send $C$ a composite error, resulting in a self-configuring 
proportional-derivative controller. 

The input correlation rule is:
\begin{equation} \label{eq:inp_corr}
    \dot{w} =  \alpha_{IC} w I_A \dot{I}_{DP},
\end{equation} 
where $I_A$ is the scaled sum of inputs from the $A$ population, $\alpha_{IC}$
is the learning rate, and $I_{DP}$ is
either $s_{DP}$ or $s_{PD}$ times a synaptic weight.

The basic rule in Eq. \ref{eq:inp_corr} was modified to avoid weights changing 
signs, and to keep
the sum of the weights constant. To make the sum of weights for connections from
$A$ to $M$ equal to $w_s$, at every simulation step we multiplied the 
weight times $\zeta_s \equiv w_s / \sum_k \omega_k$.
Weight clipping was used to keep individual weights from becoming
too large. This means that on every simulation step we set
$w = \text{min}(w, w_{max})$, where $w_{max}$ is the largest weight value
allowed.

Connections from $M$ to $C$ populations used the learning rule of equation
\ref{eq:mixed}, with the modifications of equation \ref{eq:w_norm}.

All the units (including $CE, CI$) in the architecture of figure
\ref{fig:arch_pend} were
sigmoidals as in equations \ref{eq:unit_dynamics}, \ref{eq:sig}, 
with the exception of $S_D$, $A_1$, and
$A_2$. The $S_D$ unit was a predefined function of time containing the values
that should appear in $S_P$ for a random sequence of pendulum  angles in the
range $(-0.7\pi, 0.7\pi)$. 
The $A$ population had units with a rectified logarithmic activation, modeling
sensory transducers. Their dynamics followed this equation:
\begin{equation} \label{eq:log}
        \tau_a \dot{a} = \log([1 + I -T]_+) - a,
\end{equation}
where $I$ is the scaled input sum, $T$ is a constant threshold, and $[\cdot]_+$
is the ``positive part'' function (e.g. the identity function for positive
arguments, zero for negative arguments).

The $CE,CI$ units of the pendulum controller had an additional noise term in
the dynamics of equation \ref{eq:sig}. They were integrated with the
Euler-Maruyama method. Other units were integrated with the forward Euler
method, but for the pendulum we used SciPy's (\verb+https://scipy.org/+)
explicit Runge-Kutta 5(4) method.

%

\subsection{Nonmonotonic pendulum controller}

The third plant model in this paper is the same pendulum described in 
section \ref{sub:monotonic_meth}, but the torques restricting
the pendulum's rotation were removed. 

\begin{figure}
  \centering
  \includegraphics[width=0.9\linewidth]{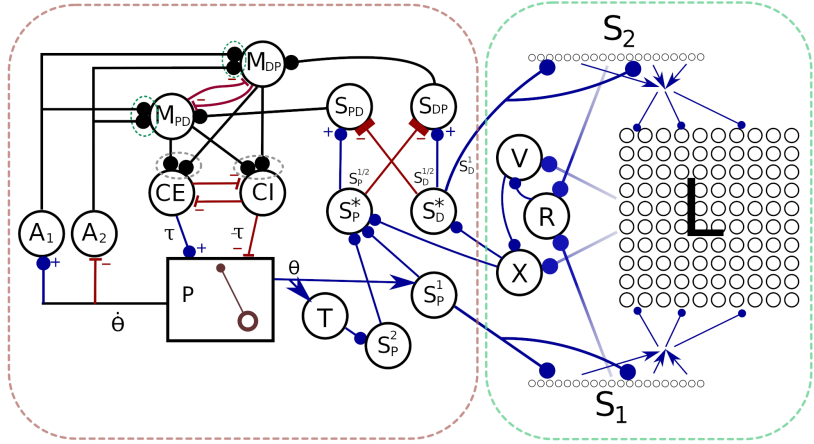}
  \caption{Actor-critic architecture used in section \ref{sub:nonmonotonic}.
  The actor component (left, red box) is similar to the feedback controller in
  figure \ref{fig:arch_pend}, but the desired and perceived angle ($S_D$ and
  $S_P$) can use one of two different coordinate systems, selected by the input
  from the unit $X$ in the critic. Moreover,
  the pendulum can rotate freely.
  The critic (right, green box) has distributed
  representations of the desired ($S_2$) and perceived ($S_1$) angles, which
  project to a state representation layer $L$. $S_1$ and $S_2$ also send
  projections to a unit $R$ that provides a reward based on the similarity of
  their activation patterns (e.g. the reward is larger when $s_1 \approx s_2$).
  $L$ sends projections to units $V$ and $X$. $V$ associates each state of the
  $L$ layer with a value, using the TD-learning rule with the reward of unit
  $R$. $X$ uses the value from $V$ to implement a version of reward-modulated
  Hebbian learning that associates each state in $L$ with an output. When the
  output of $X$ is smaller than 0.5 the actor uses a coordinate system where the
  zero degree angle lies on the positive X-axis. Conversely, when $X$'s output
  is smaller than 0.5 the actor's coordinate system has a zero degree angle
  aligned with the negative X-axis(see figure \ref{fig:coordinates}).
  The perceived angle in the coordinate system used when $X<0.5$ is
  provided by the $S_P^1$ unit. The $S_P^2$ unit provides the perceived angle in
  the alternate coordinate system, which in the simulation is obtained by having
  a unit $T$ that transforms the angle $\theta$. The $S_P^*$ unit outputs either
  $S_P^1$ or $S_P^2$ depending on the value of $X$. The $S_D^*$ units performs a
  similar function for the desired angle.
  }
\label{fig:nonmonotonic}
\end{figure}

As described in section \ref{sub:nonmonotonic}, the architecture of figure
\ref{fig:arch_pend} is limited in how well it can perform under these
conditions, but this can be improved if the controller can switch the
angle representation it uses depending on the current and desired angles.
This is done through an architecture with ``actor'' and ``critic'' components,
shown in figure \ref{fig:nonmonotonic}.

\begin{figure}
  \centering
  \includegraphics[width=0.9\linewidth]{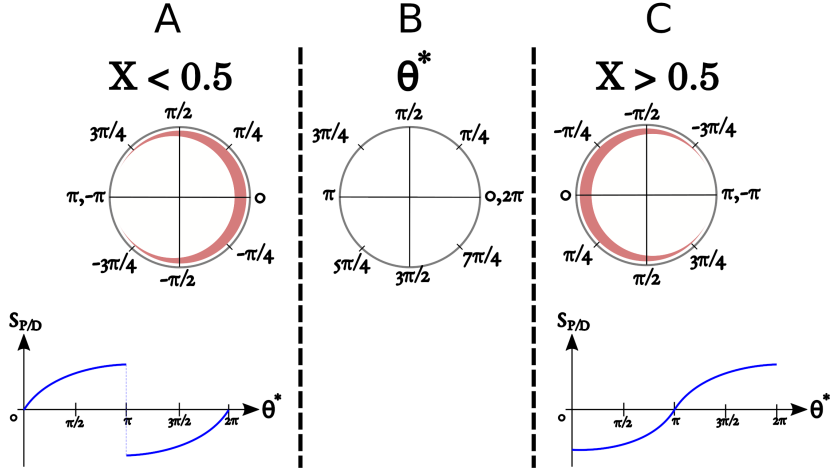}
  \caption{The two coordinate systems used in the architecture of figure
  \ref{fig:nonmonotonic}, and how they affect the activity in $S_P$ and $S_D$.
  A) Top: When the output of the $X$ unit is smaller than 0.5 the first coordinate
  system is used. In this coordinate system the plant outputs an angle in the range
  $(-\pi, \pi]$ where the zero-degrees direction is aligned with the positive
  X-axis, as shown in the circle. The thickness of the red band inside the
  circle indicates that the system can have a higher effective gain when the
  desired angle is close to zero degrees. 
  Bottom: the output of the $S_P$ and $S_D$ units as a function of the
  pendulum's angle, in the coordinate system shown in the center panel.
  B) The plots in this figure, and the angles in figure
  \ref{fig:rl_results} panels B and C are reported with respect to this
  coordinate system, where the angles are in the $[0, 2 \pi]$ range.
  C) Top: When the $X$ output is larger than 0.5 the coordinate system
  undergoes a 180-degree rotation, so that the activity of the $S_D$ and $S_P$
  units as a function of the pendulum's location is now as shown in the plot at
  the bottom of this panel.
  }
\label{fig:coordinates}
\end{figure}

The actor component in the architecture of figure \ref{fig:nonmonotonic} is the
same as the network of section \ref{sub:nonlinear}, but the torques restricting
the pendulum's rotation were removed, and additional units were introduced in
order to have an extra coordinate system that could be switched using the input
from the $X$ unit. The $T$ unit of figure \ref{fig:nonmonotonic} transforms the
$\theta$ angle provided by the plant so it uses the second coordinate system.
$S_P^1$ and $S_P^2$ are just
like the $S_P$ unit of figure \ref{fig:arch_pend}, but they differ in the
coordinate system used in their inputs ($S^1_P$ is identical to $S_P$ from figure
\ref{fig:arch_pend}, but $S_P^2$ is not). The two coordinate systems used are
described in figure \ref{fig:coordinates}.

$S_P^*$ is a unit that receives inputs from both $S^1_P$ and $S^2_P$; its output
is one of those two inputs, selected according to the value of the $X$ unit. In
engineering terms, $S_P^*$ acts like a multiplexer. When the input from $X$ is
smaller than 0.5 $S_P^*$ outputs the value from $S^1_P$, and otherwise it
outputs the value from $S^2_P$. The dynamics of $S_P^*$ follow this equation:
\begin{equation} \label{eq:sp_star}
    \tau_P \dot{s} = I - s,
\end{equation}
where $I$ is $S^1_P$ when $X<0.5$, and $S^2_P$ when $X>0.5$.
One way multiplexing can be achieved is through localized dendritic inhibition
\cite[e.g.]{jadi_location-dependent_2012}.

$S_D^*$ is the analog of $S^*_P$, providing the desired $S_P$ using one of two
possible angle representations. For simplicity, the $S_D^1$ and $S_D^2$ units
were not included in figure \ref{fig:nonmonotonic}. Desired angles were chosen
in the $(0, 2\pi)$ range.

The first component in the critic is a
distributed representation of the current perceived angle $S_P$, and desired
angle $S_D$, provided by the $S_1$, and $S_2$ populations, respectively.
$S_1$ and $S_2$ both consist of 20 units, each of
which has a bell-shaped response that increases as the input gets closer to
their preferred angle.

The information in $S_1$ and $S_2$ (and potentially other inputs) constitutes
the state or context characterizing the current situation. The state
information is combined in a single population $L$, with the purpose of
associating its activity with the right configuration for the controller. To
decide which configuration is best, we rely on reinforcement learning
techniques. In particular, $L$ provides inputs to a unit $V$ that learns a {\it
value} associated with the state using a version of the {\it temporal
differences} learning rule \cite{schultz_neural_1997}. The value provided by the
$V$ unit is used by another unit, called $X$ in figure \ref{fig:nonmonotonic}.

$X$ learns to associate the state in $L$ with an output that configures the
feedback controller. So that $X$ provides configurations that increase the
value, the connections from $L$ to $X$ use a version of reward-modulated Hebbian
learning, where the output of $V$ is used as the reward (equations 
\ref{eq:rm_hebbian1}, \ref{eq:rm_hebbian2}). 

The $V$ unit has dynamics: 
\begin{equation} \label{eq:v_dynamics}
\tau_V \dot{v} = \sigma \Big(\sum_j w^V_j L_j - \langle I_V \rangle \Big) - v,
\end{equation}
whereas the $X$ unit has dynamics: 
\begin{equation} \label{eq:x_dynamics}
\tau_X \dot{x} = \sigma \Big(\sum_j w^X_j L_j - \langle I_X \rangle \Big) - x.
\end{equation}
$\tau_V$ and $\tau_X$ are time constants,
$\sigma(\cdot)$ is the sigmoidal function, $L_j$ is the activity of the
j-th unit in $L$, and $\langle I_{V/X} \rangle$ is a low-pass
filtered version of $\sum_j w_j^{V/X} L_j$.

In the Temporal Differences (TD) learning rule \cite{sutton_reinforcement_2018}
the value function is $V(s_t) = \langle \sum_{t = 1} \gamma^{t-1} R(t) \rangle$, where 
$\gamma$ is a discount factor that reduces the importance of later versus imminent
rewards. The $V$ unit learns to approximate this function in continuous time by
adjusting its synaptic weights with the following equation:
\begin{equation} \label{eq:td}
    \dot{w_j}(t) = \alpha_{V} \big[ \bar{R} + \gamma v(t) - v(t - \Delta t_v)
    \big] L_j(t-\Delta t_v) ,
\end{equation} 
where $\bar{R} = (R(t) + R(t-\Delta t_v))/2$ approximates the integral of $R$ for
the past $\Delta t_v$ seconds. 
Two additional terms were added to this equation
in order to provide weight normalization and to have the sum of the weights near
zero. The final equation had the form: 
\begin{equation} \label{eq:add_terms}
    \dot{w_j}(t) = \Omega + \eta_1 w_j \left( \frac{W}{\sum_k |w_k|} - 1 \right) -
    \eta_2 \bar{w},
\end{equation}
where $\Omega$ is the RHS in equation \ref{eq:td}, $W$ is the desired value for
the sum of the absolute value of the weights, $\eta_1, \eta_2$
are constants, and $\bar{w}$ is the mean of all $w_j$ weights for connections
from $L$.


To adjust the weights from $L$ to $X$ we introduce a version of
reward-modulated Hebbian learning capable of handling the temporal credit
assignment problem associated with tracking a target angle in real time. For
this purpose the weights were updated intermittently, whenever the $S_D$ value
changed (e.g. whenever its derivative crossed a threshold), an event that we
will call a {\it transition}. Let $t^i$ be the time when a transition happens,
and $t^{i-1}$ be the time of the previous transition. Whether a weight is
potentiated or depressed depends on two factors. The first one is the
$V(t^i) - V(t^{i-1})$ difference, indicating whether the value increased
between transitions. The second factor is whether a sufficiently high reward
was reached, and how quickly. The concrete update equation is:
\begin{align} \label{eq:rm_hebbian1}
    \dot{w_j}(t) &= \alpha_X \Delta V(t)
    \left( L_j(t^{i-1}) - \bar{L}(t^{i-1}) \right)
    \left( X(t^{i-1}) - 0.5 \right), \\
    \label{eq:rm_hebbian2}
    \Delta V(t) &\equiv \left[ V(t) - V(t^{i-1}) + \eta_X (t-t^R) \right],
\end{align}
where $\eta_X, \alpha_X$ are constant parameters, and $t^R$ is the last time when the reward
value was above a given threshold. $t^R$ is reset after each transition. 
$\bar{L}$ denotes the average over all the $L_k$ inputs. It is
assumed that $X$ maintains a constant value between transitions, and the term
$X(t^{i-1})$ refers to the value that $X$ has in the interval $(t^{i-1},t^i)$.

The advantage of learning only at transition times
for the problem of distal rewards is discussed section \ref{sub:transition}.

Since the states in $L$ must be associated with values or configurations, it
greatly helps if the representations in $L$ are linearly separable. 
To this end $L$ does an expansive
recoding of its inputs \cite{illing_biologically_2019}
that permits $V$ and $X$ to learn functions of the state
using a single layer. The $L$ layer consisted of 100 units, arranged in a 10x10
grid. Each unit in $L$ was maximally responsive to a particular combination of
the desired and current angles, with its response decreasing exponentially 
according to the distance between the current state and its preferred angles.

The last component of the critic is the $R$ unit, which provides a reward value
based on how similar the patterns in $S_1$ and $S_2$ are. Computation of this
reward is straightforward when $S_1$ and $S_2$ have the same structure, meaning
that for each unit in $S_1$ there is a corresponding unit in $S_2$, and
vice versa. This is possible, for example, when $S_1$ and $S_2$ are two different
layers of the same cortical area, and their corresponding units are different
populations from the same microcolumn \cite{mountcastle_columnar_1997}.

The critic, as originally designed, significantly slowed the simulation.
We describe its original implementation, and how this was simplified.

In the original implementation of the critic the $S_1$ and $S_2$ populations
were units that responded maximally
when their input is close to a preferred value $I_{max}$. Their dynamics
followed the equation:
\begin{equation} \label{eq:s12}
\tau_s \dot{s} = \text{e}^{-b(I-I_{max})^2} - s,
\end{equation}
where $\tau_s$ is a time constant, $b$ controls the sharpness of the tuning, and
I is the scaled sum of inputs. The units in $L$ were sigmoidals (equations
\ref{eq:unit_dynamics}, \ref{eq:sig}), but the
connection matrices from $S_1$ and $S_2$ to $L$ ensure that each unit in $L$
responds maximally to a particular combination of $S_1$ and $S_2$ inputs. The
resulting representation is similar to radial basis functions. 

Both $S_1$ and $S_2$ had 20 units each, whereas $L$ contained 100 units.
Independently simulating the dynamics and delayed transmissions for these 140
units slowed down the simulation by an order of magnitude. Thus, for practical
reasons, the implementation of the network used multidimensional ODEs that
encapsulated the response of $L$ in a vector function. The variables in the
multidimensional ODEs do not represent the activation of the $L$ units; instead they
directly model the evolution of the synaptic weights from $L$ to $V$, and from
$L$ to $X$. The $V$ and $X$ units have consequently 101-dimensional dynamics:
100 variables for the synaptic weights, and one variable for the
output of the unit.

The activity of the $L$ ``units'' in the multidimensional ODEs was calculated
with: 
\begin{equation} \label{eq:l}
a_L = \text{e}^{-b d^2}, 
\end{equation}
where $b$ controls the
width of the tuning, and $d$ is a measure of the distance between the current
``state'', and the preferred ``state'' of the system. This ``state'' is the pair
$(\theta, \theta_D)$, containing the current and desired angle. The distance was
obtained using the $L^2$ norm, but taking into account that the angles are
periodic. 

The $V$ and $X$ units had dynamics as in equations
\ref{eq:v_dynamics} and \ref{eq:x_dynamics}, respectively.

The $R$ unit provides a reward value that indicates when the desired angle
$\theta_D$ and the current angle $\theta$ are close. This unit was implemented
as the function $r = \text{e}^{-d^2}$. Given $\theta$ and $\theta_D$ in the $[0,
2\pi]$ interval: \\
$d = \min \big(|\theta - \theta_D|, \ 2\pi - \max(\theta, \theta_D) +
\min(\theta, \theta_D) \big)$.

Learning in the connections from $L$ to $V$ used the version of TD-learning
in equations \ref{eq:td}, \ref{eq:add_terms}. 
Learning in the connections from $L$ to $X$ relied on equation
\ref{eq:rm_hebbian1}. The software implementation of this equation uses slightly
modified terms to deal with the fact that updates should happen during 
transitions (e.g. at time $t^i$), but they can't
happen instantaneously. In particular, the learning rate is modulated by a term
that decays exponentially after a transition. As with learning of the weight in
the $V$ unit, equation \ref{eq:rm_hebbian1} receives the additional terms
in equation \ref{eq:add_terms} to normalize the sum of weights and to make the
weights have zero mean.

\subsection{Inverted pendulum controller}
The fourth plant model has the same pendulum with unrestricted rotation of the
third model, but gravity is included.

The architecture used to control the pendulum is also much simpler, as described
in section \ref{sub:transition} and in figure \ref{fig:trans}.

The output of the $X$ unit approaches
either 1 or -1, depending on whether the sum of its inputs times their
synaptic weights is positive or negative, respectively:
\begin{equation} \label{eq:tanh}
\tau_X \dot{x} = \tanh \bigg( \beta \Big[ \sum_j w^X_j S_j - 
    \langle I_X \rangle \Big] \bigg) - x.
\end{equation}

$\tau_X$ is a time constant, $\beta$ is a slope parameter,
$S_j$ is the activity of the j-th unit in $S$, and $\langle I_X \rangle$ is a
low-pass filtered version of $\sum_j w_j^X L_j$.

An output of 1 produces
a positive (counterclockwise) torque $\tau$, and -1 produces a torque of
$-\tau$. $\tau$ is not sufficient to raise the pendulum from its rest position
(at $\frac{3 \pi}{2}$ radians) to an angle beyond the horizontal line. $X$ only
changes its output value at the transition times. The reward
unit $R$ has $\sin(\theta)$ as its output, providing vertical height.
The $S$ population provides a distributed representation of the angle using 20
units, in the same manner as before.

Learning in the connections from L to X relies on equation \ref{eq:rm_hebbian1}.
An additional term was used to maintain the sum of absolute weight values close
to a value $W$, leading to the equation:
\begin{equation} 
    \dot{w_j}(t) = \Omega + \alpha_X w_j \left( \frac{W}{\sum_k |w_k|} - 1
    \right),
\end{equation}
where $\Omega$ is the RHS in equation \ref{eq:rm_hebbian1}.

As described in section \ref{sub:transition},
this rule was applied at the times when $R''$ and $R'$ were negative, and
the time since the last transition was at least $t_{trans}$ seconds.

\subsection{Parameter adjustment}

Parameters for all models were manually adjusted to obtain a
reasonable dynamic range for each of the neuronal populations, 
and learning rates were adjusted
so the task could be learned relatively fast. Any other parameter
adjustments were done by trial and error, although little
parameter search was required. There were two exceptions for this.

The delays in the learning rules were obtained by an analytical procedure
described below.

The delay $\Delta t$ in the $\dot{c}_i(t - \Delta t)$ terms of the learning
rules is meant to synchronize an action in $c_i$ with the
consequent reaction in $e_j$. To this end, $\Delta t$ should contain 4
transmission delays as the signal from $C$ goes through $P$, $S_P$, $S_{DP}$,
and back to $C$. Moreover, the units at each of these stages have a response
latency. Since the equations of these units resemble those of a linear
first-order low-pass filter (e.g. Eq. \ref{eq:unit_dynamics}), 
its phase shift can be used to approximate the
response latency of the units. In particular, a signal $\sin(\upsilon t)$ 
has a filtered response $x(t)$ that is the solution of:
$\tau \dot{x} = \sin(\upsilon t) - x$. This equation can be solved exactly, and
its solution is a sinusoidal whose time delay with respect to the
input is $\arctan(\tau \upsilon)/\upsilon$. Using the most dominant frequency
observed in the activity of the units as $\upsilon$, a term like this can be
obtained for
each of the populations that the signal goes through, providing response
latencies that are added into the $\Delta t$ delay.

Parameters for the $X$ and $V$ units were first tuned manually, and then further
adjusted using 6 generations of a standard genetic algorithm, included in the
source code.

\section{Results}

\subsection{Adaptive control of a linear MIMO plant}
\label{sub:linear_control}
As described in the Methods, we produced 2 learning rules (equations
\ref{eq:rga}, \ref{eq:mixed}) to infer the input-output structure of a feedback
system. We now show how those rules performed when used to implement
proportional control of a linear plant.

As described in the Methods (section \ref{sub:linear}), the controller used the
architecture in figure \ref{fig:bio_feedback}. The plant's response came from 
a linear combination of vectors $\mathbf{v}_j$, where each vector is scaled by 
the activity of a unit in
$CE$ or $CI$. These vectors defined the connection matrix $W_{CP}$ from $C$ to
$P$, and the degree of redundancy in the controller would depend on that matrix.

We used 4 types of $W_{CP}$ matrices. $W_{CP}^{id}$ created a 
controller where each unit in $C$ affects only one error. This connection
matrix tests the simplest scenario, where the controller can act as several
independent 1-dimensional controllers; it just needs to decide which output
corresponds to which error.

The matrix $W_{CP}^{Haar}$ tests the next scenario, in which the number of units
in $CE$ (or $CI$) is equal to the dimension of the plant, but the activity of
each unit in the controller has an effect on more than one of the errors. 
The $\mathbf{v}_j$ vectors form an orthonormal basis (the Haar basis 
\cite{strang_wavelet_1993}) so in theory $C$ can produce any desired vector
output in $P$, but our system must do it by choosing the right weights in the
connections from $(S_{DP},S_{PD})$ to $(CE, CI)$. 

For the third connection matrix ($W_{CP}^{oc}$), 
the number of units in $CE$ and $CI$ is twice
the dimension of the plant. Half of the $\mathbf{v}_j$ vectors in $CE$ to $P$
connections are random unit vectors, and the other half are the
$\mathbf{v}_j$ vectors used in $W_{CP}^{Haar}$. This increases the redundancy,
not only in the sense of one controller activity $c_i$ affecting more than one
error signal $e_j$, but also in the sense that there are countless ways to
achieve a desired output in the plant.

For the final type of connectivity ($W_{CP}^{oc2}$), 
all $\mathbf{v}_j$ vectors are random, and
there are 3 for each unit in $S_P$. This is in general a much harder case, 
with greater redundancy and no guarantees of being solvable, used to illustrate
a worst-case scenario.

\begin{figure}
  \centering
  \includegraphics[width=.95\linewidth]{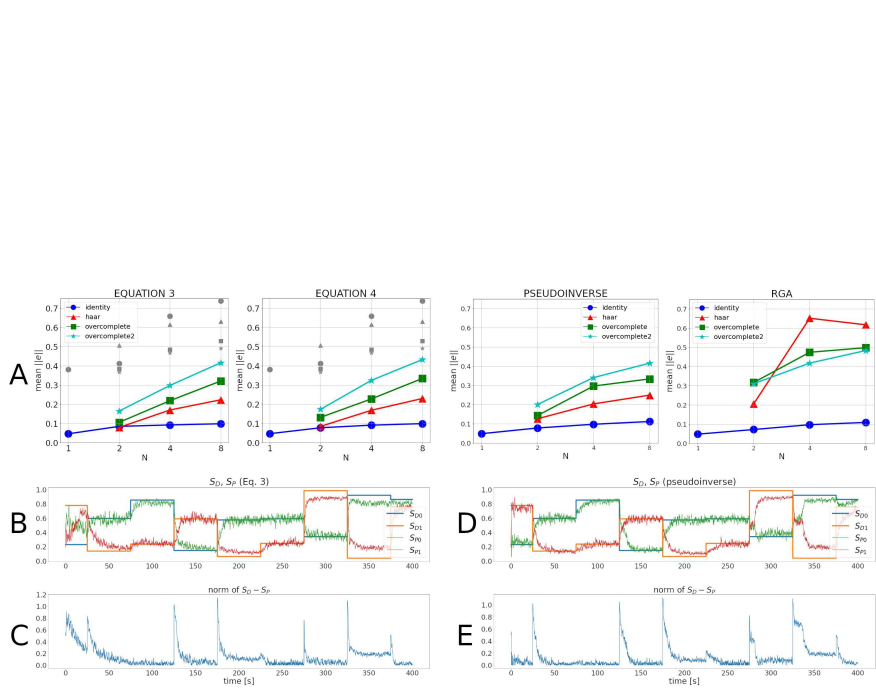}
  \caption{A) Simulation results for 4 types of connectivity matrices in a linear
  plant model for the two learning rules in section \ref{sub:rules}, and for two
  analytical methods. The number
  of values in $S_P$ is labeled $N$ in the x-axis. The y-axis indicates the
  time average of the norm $||\mathbf{s}_P - \mathbf{s}_D||$ for the second half
  of the 400 seconds simulation, where $\mathbf{s}_P$ is the vector
  of activities in $S_P$, normalized so it has a unit norm for $N>1$, and
  likewise for $\mathbf{s}_D$. Each marker is the average from 20 individual
  simulations with different random initial weights. 
  Gray markers indicate the same mean error when a
  simulation with the same characteristics was run with static synapses. In the case
  $N=1$ only the identity matrix is tested. B) Activity of the 
  $S_D$ and $S_P$ units for the first 400 seconds of an example case with N=2
  units in $S_D$ and $S_P$, an ``overcomplete'' $W_{CP}$ matrix, and the learning
  rule of equation \ref{eq:rga}, resulting in an average $||S_D-S_P||$ value of
  approximately 0.18 for the first half of the simulation, and 0.1 for the
  second half of the simulation.
  C) $||S_D-S_P||$ norm for the simulation in panel B.
  D) A simulation as in panel B, but the connection matrix from $S_{DP},S_{PD}$
  to $C$ comes from the pseudoinverse method. The $||S_D-S_P||$ average value
  was around 0.12 for both halves of the simulation.
  E) $||S_D-S_P||$ norm for the simulation in panel D.
  }
\label{fig:monotonic}
\end{figure}

Simulations are shown for 1, 2, 4, and 8 units in $S_P$, which is also the
dimension of the plant, denoted as $N$ in this section.
Results are summarized in figure \ref{fig:monotonic}. The third and fourth types of
connectivity are respectively labeled {\it overcomplete}, and 
{\it overcomplete2} in this figure.

In panel A of figure \ref{fig:monotonic} the performance of the rules is
measured as the norm of the $||S_D-S_P||$ error for the second half of the
400 seconds simulation.
The norm of the difference of two unit vectors with random
entries in the (0,1) range is expected to be around 0.5. This is a
first order approximation to the error we should expect for a system that has
done no learning. We refine this control by running simulations with random
initial weights and static synapses, resulting in the gray markers of the
first two plots.

In order to put the performance of our learning rules into context, we also
determined the input-output structure of the controller using two analytical
methods (see section \ref{subsub:analytical}). 
The first one places the Moore-Penrose pseudoinverse of the $W_{CP}$
matrix in the $W_{SP}$ matrix connecting $(S_{DP}, S_{PD})$ to $(CE, CI$). The
second one uses a simple, automated version of the RGA criterion 
\cite{bristol_new_1966}.

Quite remarkably, panel A of figure \ref{fig:monotonic} shows that the 
learning rules perform almost the same as the pseudoinverse method, and
outperform the version of the RGA method we implemented.

Both the analytical methods and the learning rules perform almost 
optimally with the system that has no redundancy (the ``identity'' case, with
the $W_{CP}^{id}$ matrix).
The error increases slightly for larger values of $N$,
because proportional control is being done in a MIMO system with delays,
response latencies, and noise, so it is inevitable that some error will 
accumulate for each controlled variable. 

The type of error that accumulates can
be observed in panels B-E of figure \ref{fig:monotonic}, showing simulation data
for the ``overcomplete'' case (with the $W_{CP}^{oc}$ connection matrix) with 
dimension $N=2$, both for the pseudoinverse method, and for the rule of
equation \ref{eq:rga}. The intrinsic noise of the $CE,CI$ units causes most of
the noisy appearance of the activity traces. Without this noise the system may 
not learn due to insufficient exploration.

In the case of $W_{CP}^{Haar}$ (red triangles in panel A of figure
\ref{fig:monotonic}), the pseudoinverse method and the two learning rules have
virtually the same performance. From here on the RGA method
largely fails, because in the simple form that we use each error is to be
controlled by a single controller unit. This is unfeasible when
each $c_i$ unit affects many $e_j$ values due to the structure of 
$W_{CP}$.

For the system with the $W_{CP}^{oc}$ connection matrix, the pseudoinverse
method and the learning rules also have similar performance. Despite redundancy,
the local rules can perform a computation that is tantamount to inverting the
connection matrix from $C$ to $P$.

In the case of the redundant, random connection matrix $W_{CP}^{oc2}$, none of
the methods performs well, as would be expected from a scenario with such level
of random redundancy.

The amount of error in the system (panels B-E) is what should be expected for
simple proportional control in this scenario. Animal motor control does not
seem to rely on one monolithic controller that does both the input-output
mapping, and ensures fast and accurate performance. Instead, there is a cerebellar
system to compensate for things such as timing, momenta, and interaction torques
\cite{manto_consensus_2012,bastian_cerebellar_1996}. Many cerebellum models 
perform this type of supplementary control (e.g.
\cite{porrill_recurrent_2004,dean_adaptive-filter_2008,kawato_computational_1992,verduzco-flores_how_2015}),
relying on a pre-existing feedback control structure.

Although these two learning
rules do not explicitly consider the full error $||\mathbf{e}||$, reducing the
components of $\mathbf{e}$ individually works well together with a type of
weight normalization that keeps the $L^1$ norm (sum of absolute values) of the
$\mathbf{e}$ vector constant. 
Normalizing incoming and outgoing weights (see Methods, section \ref{sub:rules})
allows the
network to scale its size without requiring parameter changes, and also
maintains the balance between excitation and inhibition due to the architecture
of figure \ref{fig:bio_feedback}. 

One limitation of the approach in section \ref{sub:rules} is that it requires
some knowledge of the $\Delta t$ delays inherent in the system. This is
reasonable for neurons that receive the effects of their activation with a
short, and relatively fixed latency. This would be the case, for example, of
spinal interneurons receiving feedback from muscle afferents and motor cortex.
The fact that the delay can also depend on the frequency of the oscillation (see
Methods) does not seem to impair the system, as only few dominant frequencies
tend to naturally emerge.

\subsection{Monotonic control of a pendulum}
\label{sub:nonlinear}
The linear plants in section \ref{sub:linear_control} show how that the
learning rules can resolve moderate amounts of redundancy in the controller,
but they are not representative of physical systems. Next we consider
feedback control of a pendulum.

The error signal in this case is the difference between desired and current
angles. So that this error remains monotonic we make the pendulum stop when
it approaches $\pm\pi$ radians (see Methods). This, however, does not change
the fact that simple proportional control (as in the
architecture of figure \ref{fig:bio_feedback}) may be unstable, despite the
addition of viscous friction. This is due to the delay
in the control response, which is similar to the delays observed in human
reflexes \cite{capadayReexaminationEffectsInstruction1994}. Such an effect
highlights the usefulness of including transmission delays and response
latencies in this study.

As discussed previously,
most cerebellar models assume a pre-existing feedback controller,
whose performance they improve. And as discussed in section 
\ref{sub:dividends},
configuration of this feedback controller may not be innate. If this is the
case, the feedback controller can't rely on the cerebellum while it is learning
its input-output structure, and must somehow compensate for its unstability.

In systems where proportional control is unstable, oftentimes
proportional-derivative control can restore stability
\cite{sontag_mathematical_2013}. Animals can receive muscle contraction velocity
and tension information from their muscle afferents
\cite{shadmehr_computational_2005}. We extended the architecture of figure
\ref{fig:bio_feedback} to include angular velocity information while still 
allowing for
self-configuration using the learning rules of section \ref{sub:rules}.
The result is the architecture in figure \ref{fig:arch_pend}.

The $S_D$ population in figure \ref{fig:arch_pend} does not specify 
a desired velocity, so a velocity error
cannot be produced in the same way as the angle error. In order to
adaptively incorporate the velocity information into the control loop we
created a network resembling the
long-loop reflex of the animal motor system, which includes not only the spinal cord,
but also the primary motor and sensory cortices. 

In Figure \ref{fig:arch_pend} we introduced a population $M$ receiving the 
afferent activity $A$, consisting of the angular velocity $\dot{\theta}$ in its 
non-negative (dual) representation. In addition, each $M$ unit received one
error signal, either $s_{DP}$, or $s_{PD}$. The $M$ units used the
\emph{input correlation} rule \cite{porr_strongly_2006} (equation
\ref{eq:inp_corr}) to potentiate
angular velocity inputs that correlate with their error input.
This allows $M$ to send $C$ a composite error, resulting in a self-configuring 
proportional-derivative controller. 

$C$ uses two units, one providing clockwise, and another counterclockwise torque.
The value in $S_D$ represents a given angle, and the task is to move the pendulum
to that angle so activity in $S_P$ and $S_D$ can be equal. 

\begin{figure}
  \centering
  \includegraphics[width=1.0\linewidth]{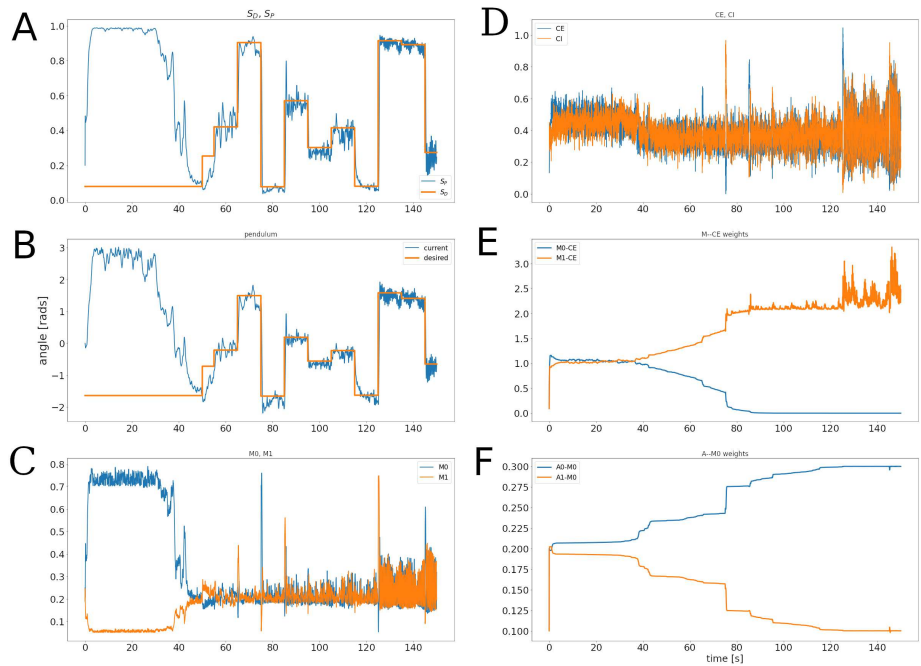}
  \caption{ First 150 seconds of a simulation where the architecture of figure
  \ref{fig:arch_pend} is used so a pendulum can track a desired angle
  (no gravity). The system learns to track the desired angle in
  about 60 seconds.
  A) Activity of the $S_P$ unit, with the perceived angle, and $S_D$, with the
  desired value for $S_P$. B) Angle of the pendulum, and the desired angle.
  C) Activities of the two units in population $M$. D)
  Activities of the two units in population $C$. E) Synaptic weights for the
  connections from the two $M$ units to the $CE$ unit. F) Synaptic weights for
  the connections from the two $A$ units to one of the $M$ units.
  }
\label{fig:pendulum_1}
\end{figure}

Figure \ref{fig:pendulum_1} shows a representative simulation result, where the
system learns to perceive a desired $S_D$ angle in $S_P$ using the learning rule from
equation \ref{eq:mixed} in a pendulum with no gravity. A similar figure for the 
case when gravity is present is in \ref{app:gravity} (figure
\ref{fig:pendulum_2}). Figure \ref{fig:pendulum_1} shows the 
appropriate weights emerging in seconds; this time depends on the initial 
conditions and the learning rates. After a couple of
minutes the weights reach their final values, which remain stable thereafter.

An interesting feature of this system is the interplay between antagonist (dual) 
units, seeking a balance between excitation and inhibition. Panel B of figure 
\ref{fig:pendulum_1} shows how each time the target changes one of the $M$ units
activates more than its dual, producing a correction. The magnitude of the 
error determines difference in the activity of dual $M$ units.
In the absence of
gravity the error can remain close to zero without exerting any torque, and at
this equilibrium point both $M$ units have the same activation level, sending no
net excitation to $CE$ and $CI$. 

All the units in figure \ref{fig:arch_pend} have a sigmoidal activation
function, except for those in population $A$, which have a logarithmic
activation (equation \ref{eq:log}). Sigmoidals have a 
non-zero output in the absence of input (equation \ref{eq:sig}). Thus, in
the absence of error the units may still have an output, but antagonist units
will have the same activation level, resulting in no action. When gravity is present
a constant torque is required to keep the error close to zero. Since the system
exerts no action in the absence of error, gravity  implies that either we will
have a steady state with non-zero error, or the angles
will oscillate around their target values. Which of these scenarios presents
depends on the gain of the system, with higher gains tending to produce
oscillations around the target. Moreover, the $C$ units present intrinsic noise,
used so the system can produce plasticity-inducing movements when learning
begins. All of these factors explain the oscillations observed in the figures
\ref{fig:pendulum_1} and \ref{fig:pendulum_2}. 

\subsection{Non-monotonic control of a pendulum}
\label{sub:nonmonotonic}
The two terms in the synaptic learning rules of equations 
\ref{eq:rga} and \ref{eq:mixed} are 
monotonic functions of $\dot{e}$ (or $\ddot{e}_j$) and $\dot{c}_i$.
If $c_i$ activity can make $e_j$ either grow or decrease
depending on the context, correlations will be inconsistent, making the
approach used by these equations unlikely to succeed.

A further complication is that the representation of sensory signals may not
always be germane for negative feedback control. Muscle afferents use a firing
rate code that provides information about the muscle's length, speed, and
tension, but other afferents may provide a distributed representation, using a
population of neurons where each one is tuned to a particular range of values
(e.g. direction tuning in somatosensory cortex \cite{pei_shape_2010}, 
or retinotopic location tuning in posterior parietal cortex 
\cite{andersen_encoding_1985}).

It is evident that learning a static input-output structure for a feedback 
controller is not sufficient for the control of arbitrary plants. Much
flexibility could be gained if the input-output structure could adapt according
to the context. To this end, we borrow concepts from the {\it
actor-critic architecture} used in reinforcement learning 
\cite{sutton_reinforcement_2018}. The general idea is to have a feedback
controller as an {\it actor} component that can adapt its input-output structure.
When entering a context where the current controller structure is not
appropriate, a {\it critic} component can indicate this, so the controller
alters its {\it configuration}.

The meaning of ``altering the controller configuration'' can have several
interpretations (see Discussion, section \ref{sub:hierarchies}). We present one
illustrative example in this section.

Consider the architecture in figure \ref{fig:arch_pend}, and suppose the
pendulum was able to rotate without restrictions.
Given our choice of angle representation in the $S_P$ and $S_D$ units (selected
to mimic the representation of length and velocity used in muscle afferents),
letting the pendulum rotate freely will produce a discontinuity around
$\pi$ radians, where a small variation in the angle
creates a large variation in the firing rate. 
This simple change greatly alters the pendulum control problem from the
previous section, in the sense that an optimal solution can no longer be
achieved by a controller that responds proportionally to $(\theta_D - \theta$),
where $\theta_D$ is a desired angle, and $\theta$ is the current angle.
This is because the proportional controller will not cross the angle where it
has a representation discontinuity, so even if $\theta_D$ and $\theta$ are very
close (say, 179 degrees and 181 degrees), the controller may not move the
pendulum through the shortest path. An optimal solution is thus beyond the reach
of the learning rules in section \ref{sub:rules}, which cannot handle the
non-monotonicity present in the angle discontinuity.

Because of this phenomenon we can test our ideas directly on the
pendulum controller of the last
section, with minimal modifications. In particular, we allow the pendulum to
rotate freely, but we also add the possibility of using a
different angle representation (inspired by how corticospinal signals can
modulate ascending afferents through presynaptic inhibition 
\cite{goulding_inhibition_2014}). In this way the synaptic learning rules from
section \ref{sub:rules} can still be used as before. We also add a ``critic''
component to the architecture, used to select which angle representation is
used. The result is shown in figure \ref{fig:nonmonotonic}, and details are
in the Methods section.

In abstract terms, the ``critic'' has a
representation of the state, including the desired and perceived 
angles for the controllers. From this, it produces a value associated
with each state, and this value is used to configure the controller, which in
this case means selecting an angle representation (figure \ref{fig:coordinates}).

Allowing the critic to select the coordinate system for
each state significantly increases the average value of the reward (the
output of the $R$ unit) in the case where the gain of the inputs from
$CE$ and $CI$ to $P$ is reduced.
Optimal performance in this task has to leverage two limitations. First, as
mentioned above, when the shortest path between the current and desired angles
crosses either 0 or $\pi$ radians, one of the angle representations makes
the controller follow the longer path. This affects the time to approach the 
desired angle. Second,
due to the limited dynamic range of the sigmoidal units, the gain of the
controller is greatly reduced when the desired angle is away from the
zero-degree direction (figure \ref{fig:coordinates}). The critic must thus
choose a coordinate system that has enough gain near the desired angle. This
affects the error in the steady state.

\begin{figure}
  \centering
  \includegraphics[width=1.1\linewidth]{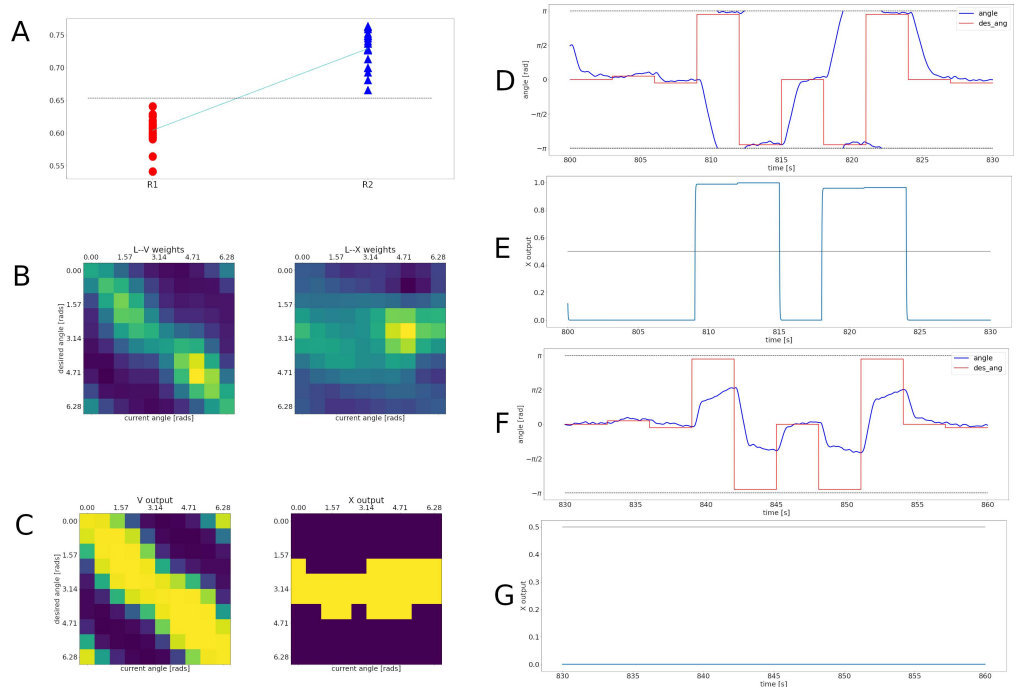}
  \caption{Performance of the reinforcement learning model. A) average reward
  in 20 simulations when the $X$ value is randomly selected (R1, red circles)
  compared to the reward when the $X$ value is produced by the inputs from $L$
  after training (R2, blue triangles). 
  B) Left: Connection weights for the projections from $L$ to $V$ after
  training. These weights were selected arbitrarily from one of the 20
  simulations used for panel A. Right:
  Connection weights for the projections from $L$ to $X$ after training.
  C) Left: Steady-state activation values for the $V$ unit when the desired and
  current angles are those preferred by each of the 100 units in $L$. Right: the
  corresponding steady-state activation values for $X$.
  D) The desired (red) and current (blue) angles through a 30 second simulation.
  This panel shows tracking of the desired angles after 800 seconds of learning
  with random $X$ values. After this learning period the $X$ values
  were determined by the input from $L$, and the 30-second simulation in
  this plot began.
  The desired angles were selected to illustrate the difference when using the
  actor-critic system, compared to simple feedback control as in section
  \ref{sub:nonlinear}. Angles in the y-axis are in the coordinate
  system of panel A in figure \ref{fig:coordinates}.
  E) The output of $X$ during the 30-second simulation of panel D.
  F) A simulation with the same desired values as in  panel D, but this time the
  output of $X$ was fixed near 0, forcing the use of a single coordinate
  system.
  G) The output of $X$ during the 30 seconds of the simulation in panel F.
  }
\label{fig:rl_results}
\end{figure}

Figure \ref{fig:rl_results} shows the results of 20
simulations where the network was first run for 800 seconds
with random $X$
values (either $X \approx 0$ or $X \approx 1$ on each reach) to provide a mean
reward $R1$. Next the network was run for 400 seconds with the
$X$ output being
driven by the inputs from $L$, providing a mean reward $R2$. The average
increase in reward was approximately 0.136 ($p < 1^{-10}$, paired T-test), and the
largest $R1$ value in the 20 simulations was smaller than the smallest $R2$ 
value.

In the simulations presented in figure \ref{fig:rl_results}  a different $S_D$
value was presented every 4 seconds. In the first 800 seconds the feedback
controller would attempt to make $S_D = S_P$ using one of the two angle 
representations, selected randomly, and as it did so learning took place in the
connections from $L$ to $V$ and from $L$ to $X$. The
$V$ unit was learning to estimate the value of different states, and the $X$
unit was learning which output was associated with an increase in this value.

The strategy that emerged through learning can be glimpsed from the weights in
the projections from $L$ to $X$, and the outputs that they implied, as shown in
panels B and C of figure \ref{fig:rl_results}.
In this figure the horizontal axis represents
the current angle, and the vertical axis represents the desired angle. Each
of the squares in this 10x10 grid correspond to the unit in $L$ that is
maximally responsive to the corresponding combination of angles.
In panel B the color of the square encodes the magnitude of the synaptic weight
in the projection of that $L$ unit to $X$, with brighter squares having a larger
weight. In panel C (right half) yellow squares indicate an output close to 1, which
causes the second coordinate system to be used (figure \ref{fig:coordinates}C).
As can be observed, this second coordinate system is preferred when the
desired angle is close to $\pi$ radians, whereas the first coordinate system is
preferred when the desired angle is close to 0 or $2\pi$ radians.
The effect that this has on the tracking performance can be observed by
contrasting panels D and F.

\subsection{Control of an inverted pendulum}
\label{sub:transition}
In the actor critic architecture of section \ref{sub:nonmonotonic}
the weights of the unit $X$ are updated when the desired angle changes. 
We refer to these events as {\it transitions}. Let $t_1$ denote the time
when a transition happens, and let $t_0$ be the time when the previous
transition occurred. The weight update rule (equation \ref{eq:rm_hebbian1}) only
cares about the difference in values $V(t_1) - V(t_0)$, with a possible time
penalization to discourage large ($t_1 - t_0)$ periods. Ignoring the
intermediate $V(t)$ values allows the controller to explore the gradient of the
value function in larger steps. In this subsection we present a simple
example to illustrate how this idea can be exploited.

\begin{figure}
  \centering
  \includegraphics[width=0.9\linewidth]{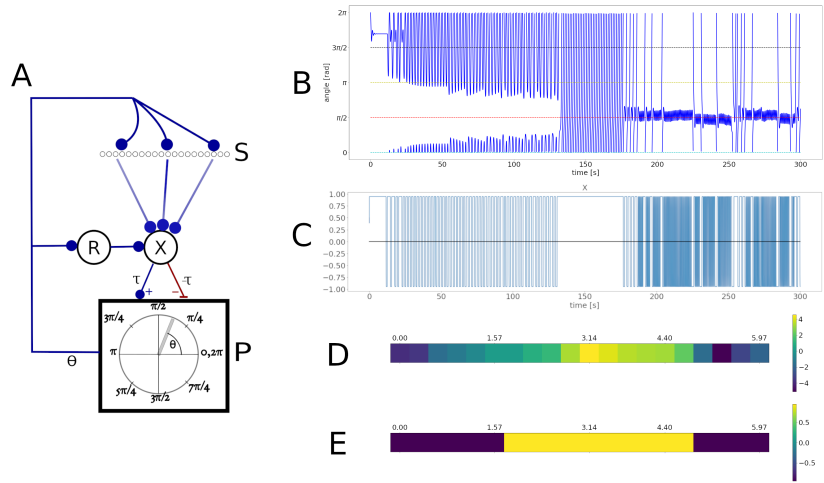}
  \caption{An architecture for the inverted pendulum problem, and simulation
  results. A) The network consists of a population $S$, plus units $X$, and $R$.
  $S$ represents the angle of the pendulum using 20 units, each with a preferred
  angle. Unit $R$ outputs the sine of the angle, which is received by unit $X$
  as a reward value. Unit $X$ uses reward-modulated Hebbian learning (equation
  \ref{eq:rm_hebbian1}) with the right update times to adjust the weights in the
  connections from the population $S$.
  B) The pendulum's angle through the first 300 seconds of a simulation.
  Initially the pendulum oscillates at the bottom, around the $3\pi/2$ angle
  (black dotted line). Eventually the pendulum manages to complete a revolution,
  and begins to spin, until it starts to balance at the top, around the $\pi/2$
  angle (red dotted line).
  C) Output of the $X$ unit, which is proportional to the torque applied.
  D) Weights in the connections from $S$ to $X$ after 300 seconds. Each square
  corresponds to the weight of a particular unit in $S$, and the label at the
  top indicates the preferred angle of that unit.
  E) Steady-state output of $X$ when the angle is one of the preferred angles of
  the 20 units in $S$.
  }
\label{fig:trans}
\end{figure}

Consider the {\it inverted pendulum} problem, where the goal is to make the
pendulum reach the vertical position, at $\frac{\pi}{2}$ radians in the coordinate
system of panel B in figure \ref{fig:coordinates}. This problem is
trivial using a controller as in the previous subsections, with enough gain to
overcome gravity. To make this example illustrative we removed the controller
and most of the critic from the architecture of figure \ref{fig:nonmonotonic},
leading to the reduced system in figure \ref{fig:trans}A.

Learning in this model happens in the connections from $S$ to $X$, using the
reward-modulated Hebbian rule of equations \ref{eq:rm_hebbian1},
\ref{eq:rm_hebbian2}. This system will generally
not learn to point the pendulum upwards using random transition times; it is
necessary to have a particular strategy. Denoting the output of $X$ as the {\it
configuration}, we outline our strategy as follows:

\begin{enumerate}
\item Adopt a configuration (e.g. give $X$ a fixed output
value).
\item Predict the time $t^*$ when $V(t)$ will attain its maximum 
      (updating the prediction online).
\item Perform a transition at time $t=t^*$.
\end{enumerate}

The inverted pendulum problem is simple enough that a value function $V$
and a controller $C$ as those in figure \ref{fig:nonmonotonic} are not
required. In the case of figure \ref{fig:trans} $R$ takes the place of $V$,
and $X$ takes the place of $C$. For this particular case the strategy above can
be adapted into a simple rule: {\it if both $R''<0$ and $R'<0$, do a transition
every $t_{trans}$ seconds.}

This rule comes from estimating $V(t)$ (in our case, $R(t)$) as a quadratic
polynomial function of time: $V(t) = V''(t)t^2 + V'(t)t + V(t_0)$, using the
latest observed values of $V''(t)$ and $V'(t)$. If we want to maximize $V(t)$,
having $V''>0$ means that eventually the value will grow as time increases, so
no transition should be made. On the other hand, if $V''<0$, the polynomial
attains its maximum value at the point when $V'$ becomes negative, so no
transition should be done while $V'>0$. The parameter $t_{trans}$ (which could be
a random value) determines how much time the controller is allowed to explore a
configuration before a different one is potentially adopted.

The result of using this rule to decide when to apply weight updates with
equation \ref{eq:rm_hebbian1} is shown in figure \ref{fig:trans}.
The controller only has two possible torques, the angle
representation is not very precise, and there are temporal delays, so the best
that can be expected is oscillations near the $\pi/2$ angle. 
Still, the system learns to maintain the pendulum near the vertical
position for extended periods of time, and it brings it back on top soon after
it falls (figure \ref{fig:trans}B).

\section{Discussion}

\subsection{From correlations to reinforcement learning}
In this paper we presented synaptic learning rules that
automatically configure a feedback control system. This control system is
entirely agnostic about the plant being controlled, so its configuration
involves finding the input-output structure of the controller, which is akin 
to finding the sensitivity derivatives, or the control Jacobian of the system.

In section \ref{sub:rules} we derived 2 different learning rules to find this
input-output structure in the case of a monotonic relation between the control
signals and the error, and 2 other variations are in \ref{app:alternative}.
The basic form of those four equations can be written as:
\begin{equation} \label{eq:general}
        \dot{\omega}_{ij} = -\alpha \Psi(\mathbf{e}(t))\Gamma_j(\mathbf{e}(t))
        H_i(\mathbf{c}(t)),
\end{equation}
where $\alpha$ is a learning rate, $\Psi(\mathbf{e}(t))$ is an operator to  
measure the error gradient, $\Gamma_j(\mathbf{e}(t))$
quantifies the input activity, and $H_i(\mathbf{c}(t))$ quantifies the
postsynaptic activity. In the case of equations \ref{eq:rga}, and \ref{eq:mixed}
we have $\Psi = 1$ because the input is an error, and the term
$\Gamma_j$ can play the parts of both error gradient and input activity.

From this optic, the rules in this paper are not far from previous forms of
node perturbation \cite{mazzoni_more_1991,williams_simple_1992}, 
reward modulated Hebbian learning (e.g
\cite{legenstein_reward-modulated_2010,fremaux_functional_2010}), or Hebbian descent 
\cite{melchior_hebbian-descent_2019}.
We went beyond previous approaches in order to deal with complications
from continuous-time control with delays. This required using other elements
like derivatives, time delays, and normalization.

The rules in section
\ref{sub:rules} and \ref{app:alternative}
have two obvious drawbacks. One is that the error gradients do not
take distal outcomes into account. In other words, 
the learning rules can only reduce errors that
happen soon afterwards (on the order of $\Delta t$), but errors that happen
later cannot be preemptively corrected.
The second drawback is the restriction to monotonic control, since
equation \ref{eq:general} has no context information beyond the $\mathbf{e}$ and
$\mathbf{c}$ vectors.

Under this perspective, the actor-critic architecture in section 
\ref{sub:nonmonotonic}, through equation \ref{eq:rm_hebbian1}, improves over
equation \ref{eq:general} by providing modulation that can handle temporal
credit assignment, and can consider a more general context in order to produce
an output.

Learning in equation \ref{eq:rm_hebbian1} can solve the temporal credit assignment
problem because of two traits. The first trait is the use of a
value function, which considers future rewards when the discount factor is not
zero. The second trait is that updates
are performed intermittently, only during the transition times. This allows to
flexibly span arbitrary lengths of time, but it opens
the question of when should the transitions happen. We began to
address this question in section \ref{sub:transition}.


Equation \ref{eq:rm_hebbian1} can handle general context dependencies because
the state information is present in layer $L$, which uses an expansive recoding
so that $X$ can approximate arbitrary functions of the state.
A possible problem with expansive recoding is that the
number of units required scales poorly with the dimension of the input. Other
possibilities could include special versions of self-organizing maps
\cite{kohonen_variants_1995}, or a more biological version of the state
representations used in deep reinforcement learning. Notice also
that the layer $L$ is reminiscent of the sensory maps used in direct inverse
learning \cite{kuperstein_neural_1988} to associate afferent inputs with
muscle activities. $L$ could be seen as a more general version of these maps,
also representing desired values, and not necessarily being used to produce muscle
activations, but control signals at a higher hierarchical level.

\subsection{Hierarchies of feedback controllers}
\label{sub:hierarchies}
As mentioned in the Introduction, a promising idea on how to generate flexible
motor control is to have a hierarchy of feedback controllers that ultimately
regulate the value of homeostatic variables for the organism
\cite{powers_behavior:_2005}. A clear complication is that higher levels
of the sensorimotor hierarchy may deal with abstract representations, where an
error cannot be obtained by a mere subtraction operation.
The architecture we have introduced in section \ref{sub:nonmonotonic}
may open the path to exert feedback control with complex
representations.

\begin{figure}
  \centering
  \includegraphics[width=0.9\linewidth]{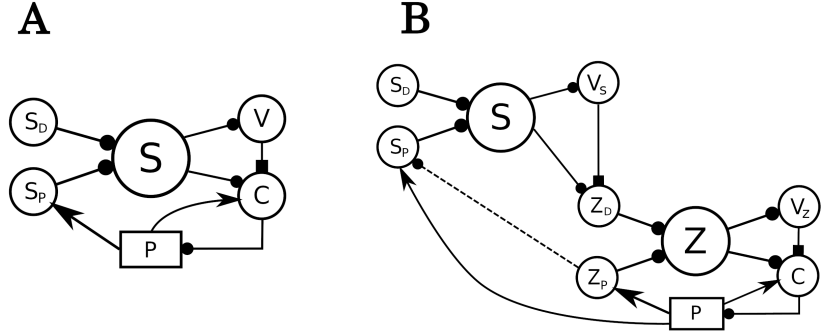}
  \caption{A) A reinterpretation of the architecture of figure
  \ref{fig:nonmonotonic} as a 2-level hierarchical control system. Arrowheads
  denote afferent connections, squares modulatory connections, and
  circles all other synaptic connections.
  The population $C$ is considered as a final controller, possibly in the spinal
  cord. The loop from $C$ to $P$ and back represents a
  first-level controller using an error representation amenable to
  negative feedback control. The level on top of this provides the ability to
  use a distributed representation for the desired and perceived values.
  B) A 3-level hierarchy of feedback controllers.  A high-level desired
  perception $S_D$, together with the current perception $S_P$ are expanded into
  a high-level state $S$, which is used to produce a value $V_S$. This value,
  the state $S$, and the current perception $S_P$ can potentially be used to
  configure a controller lower in the hierarchy, whose target value is expressed
  by the $Z_D$ population. The dotted connection from $Z_P$ to $S_P$ expresses
  that the representation in $S_P$ could be constructed using lower level
  representations rather than state variables of the plant.
  }
\label{fig:hierarchy}
\end{figure}

The basic idea of a general feedback controller can be explained with the
diagram in panel A
of figure \ref{fig:hierarchy}. $S_P$ and $S_D$ can use arbitrary
distributed representations, but because these two layer have the same structure
we can always detect when their activity is very similar, an event that would
produce the reward signal used by $V$ to learn. All of the relevant state
information is present in a population $S$, and this is used by $V$, as well as
by the controller $C$. The $C$ circle in figure \ref{fig:hierarchy} is not a
unit; it encompasses a feedback controller, and the elements that allow its
configuration. In the case of the architecture of figure \ref{fig:nonmonotonic}
this would include the actor and the $X$ unit.  
$C$ uses the value provided by $V$ in order to learn its configuration,
and the information in $S$ in order to perceive the state.  

In our example we set $S_D$ to be the desired activation caused by the target
angle in the controller $C$, but other things could be encoded in $S_D$,
such as the target in a
different coordinate system for a more complex controller. The network
comprising the $S_D$, $S_P$, $S$, and $V$ populations could be considered as a
separate control system, where $V$ provides a measure of the distance in the
activities of $S_D$ and $S_P$, and this is used either to configure, or to set
the target value of the controller $C$. Learning happens in stages, where the
lower-level controllers learn first, and the higher-level controllers perform
significant learning after the lower levels can match their target values. In
the example of section \ref{sub:nonmonotonic} the feedback controller is already
operating while the reinforcement learning system refines its operation, a
trait that should be useful for biological organisms.

Configuration of a controller using a value function can have several
interpretations. In the example of section \ref{sub:nonmonotonic} this meant
selecting the afferent input. Alternatively, this could mean selecting a
different controller altogether, which would provide a different implementation
of ideas in the MOSAIC-MR model \cite{sugimoto_mosaic_2011}, where different RL
controllers are
used depending on the context. Controller selection has also been suggested as
the main role of the basal ganglia \cite{yin_role_2006,yin_basal_2017}.

Most interestingly, the architecture of figure \ref{fig:nonmonotonic}, being a
feedback controller that configures a feedback controller, naturally has a
hierarchical extension, shown in the panel B of figure \ref{fig:hierarchy}. 
A high-level controller with ``$S$'' populations is used to configure a lower
level controller with ``$Z$'' populations, possibly setting the desired value
$Z_D$. Transforming the pair $S_D, S_P$ into a $Z_D$ value is akin to a
coordinate transformation, but in this setting it can also be conceived as
a process of subgoal selection. By generating rewards for level 
$S$ when a ``$\mathbf{s}_P = \mathbf{s}_D$'' event occurs we can learn a value
function for the $V_S$ unit. The output of $V_S$ can be used to modulate
plasticity in the descending connections from the $S$ level to the $Z$ level.
This last level receives rewards when the ``$\mathbf{z}_P = \mathbf{z}_D$''
event happens.
Having a natural reward function at each level, and
the ability to deal with distal rewards gives the model the potential of
tackling the problem of finding subgoals, which is common in the hierarchical
reinforcement learning literature (e.g. 
\cite{vezhnevets_feudal_2017,kulkarni_hierarchical_2016}).

One promising idea is to create sensory representations by grouping states 
that succeed with similar controller configurations. A direction of future 
research is thus to use the hierarchical architecture of this model to test 
whether this controllability criterion can facilitate the formation of 
perceptual categories.

\subsection{The dividends of biological plausibility}
\label{sub:dividends}
Our model suggests a coherent set of hypotheses
regarding animal motor control. We outline this below.

\begin{itemize}
  \item Spinal cord plasticity, and how it coordinates with
  plasticity at other cortical and subcortical sites
  is a challenging issue (\cite{wolpaw_complex_1997,norton_acquisition_2018,
  brumleySpinalCordNot2018}).
  Plasticity rules like those of section \ref{sub:rules}, if present in the
  spinal cord, could enable it to become a self-configuring feedback controller.
  This idea has been suggested before 
  \cite{raphaelSpinalLikeRegulatorFacilitates2010}, but a plausible plasticity
  mechanism has been missing. Furthermore, the model of section
  \ref{sub:nonmonotonic} shows how plasticity at four different sites can
  coordinate in a hierarchical manner.

  \item Some motor control models, such as feedback error learning 
  \cite{miyamoto_feedback-error-learning_1988}, posit that knowledge about
  sensitivity derivatives $\frac{de}{dc}$ is innate, rather than learned.
  However, there is significant evidence that some systems recover when the
  relation between motor command and error is reversed, so $\frac{de}{dc}$
  changes sign \cite{lillicrapAdaptingInversionVisual2013,
  sekiyamaBodyImageVisuomotor2000, richterLongtermAdaptationPrisminduced2002,
  yamashitaRestorationContralateralRepresentation2012,
  abdelghaniLearningCourseAdjustments2010,
  sachseWorldUpsideInnsbruck2017, kuangWhenAdaptiveControl2015}.
  Our model is consistent with this, and it further predicts that
  in some cases animals may be able learn to use opposite estimates of
  $\frac{de}{dc}$ depending on the context, but this learning should be much
  slower, as it depends on a reinforcement learning mechanism
  (cf. \cite{lillicrapAdaptingInversionVisual2013}).
   


  \item Feedback control is naturally limited by response latencies,
  and gains that saturate, so a cerebellar module to improve performance is an
  ideal complement. We emphasize 3 facts: 1) the cerebellum is involved in
  estimating the timing of events \cite{baresConsensusPaperDecoding2019},
  2) the cerebellum contains predictive signals in the scale of tens of milliseconds 
  \cite{tseng_sensory_2007,ebner_cerebellum_2013,baresConsensusPaperDecoding2019,
  herzfeldEncodingActionPurkinje2015,tanakaCerebroCerebellumLocusForward2020}, and 
  3) disynaptic or monosynaptic projections from the cerebellum can be 
  found in spinal cord, as
  well as cerebral cortex and basal ganglia 
  \cite{middletonChapter32Dentate1997,bostanBasalGangliaCerebellum2018,
  liangProjectionsBrainSpinal2011,nudoDescendingPathwaysSpinal1989}.
  If the cerebellum relays
  signals anticipating events at the spinal cord, they could be inputs to the $C$
  units in our model, and learning would enable the spinal controller to use
  these signals to drive anticipated responses. An inverse model in the
  cerebellum is thus not required for this type of adaptation. On the other hand,
  supraspinal projections from the cerebellum could be involved in signaling
  transition times when particular events are anticipated. We thus hypothesize
  that cerebellar signals to the spinal cord can drive anticipatory responses,
  and that signals to the cortex and basal ganglia can change the timing of
  reward-modulated plasticity.

\item Using only positive activations and weights that do not change sign
  motivates the use of dual representations, where the excitation in one neural
  population caused by a sensorimotor event should come together with inhibition
  in another population. This is not only consistent with experimental
  observations (e.g. \cite{shafi_variability_2007,steinmetz_distributed_2019,
  najafiExcitatoryInhibitorySubnetworks2020}),
  but it also permits the function of learning rules as
  the ones in section \ref{sub:rules}. 
  When controllable signals exist in antagonistic pairs, it is natural that
  the activity of a unit does not necessarily produce
  an action; what matters is the balance between excitation and inhibition.
  Balance between excitation and inhibition (E/I balance) has received extensive
  experimental validation, and has been largely recognized as
  necessary for theoretical models to reproduce observed neuronal dynamics
  \cite{dehghaniDynamicBalanceExcitation2016,haiderNeocorticalNetworkActivity2006,
  okunBalanceExcitationInhibition2009,
  okunInstantaneousCorrelationExcitation2008}.
  Our framework explains why concomitant excitatory and inhibitory responses to
  sensory events should be prevalent, and links it to the E/I balance using a
  functional model.
  

\end{itemize}

We took all these insights into a more comprehensive model of mammalian arm
reaching, where the complexity of the plant and the biological realism of the
controller were enhanced \cite{verduzco-flores_adaptive_2021}. While the
detailed findings of this follow-up paper are outside the scope of this work,
we can briefly mention that using the learning rule in equation \ref{eq:mixed}
as a self-configuration mechanism for signals in the spinal cord, we can produce
2D reaching from scratch, and explain the emergence of directional tuning in
motor cortex, among other phenomena.

\subsection{Comparison with previous work}

As mentioned in the Introduction, the closest approach to our
work is in \cite{abdelghani_sensitivity_2008}. This work required a separate
network, and represented the sensitivity derivatives using firing rates. It does
not address delays or response latencies, was implemented in discrete time
steps, and also controls simple systems (the vestibulo-ocular reflex, and the 
forearm angle of a 2-joint arm). The learning times are similar to our models.

The review in \cite{kolodziejski_mathematical_2008}
describes learning rules working in simple open-loop
circuits. Two of these learning rules could potentially be compared to our own,
namely the ISO \cite{porr_iso_2003}, and the ICO \cite{porr_strongly_2006}
rules. When applied to control problems, both rules begin by assuming that there
is an already established feedback control system whose performance is hindered
by response delays. Both rules can autonomously learn how to improve the
system's performance by applying predictive feedforward responses. The system is
thus not learning sensitivity derivatives, or in other words, it does not learn
which control signals are capable of reducing particular errors in the MIMO
closed-loop setting, which is what our rules achieve. The ISO and ICO rules
could thus be used in conjunction with our rules, which would be used to
configure the underlying feedback control system.

A similar observation applies to work based on feedback-error learning
\cite{kawato_computational_1992}, and on the recurrent architecture
\cite{ porrill_recurrent_2004} as they rely on a previously existing feedback
controller whose output is used to train an inverse model. This feedback
controller must already have the right input-output structure, or learning will
fail. Finding this input-output structure can be done by our learning rules when
this is not explicitly specified.

The distal learning approach of \cite{jordan_forward_1992} does have the
potential to fully perform controller configuration, but this relies on
backpropagating an error signal through a forward model, which strains
biological plausibility.

There is also a relatively large number of neurobiomechanical models that
perform simple motor tasks. In general they are not relevant here due to one
or more of the following reasons:
\begin{enumerate}
\item They do not address the problem of input-output configuration (e.g.
finding sensitivity derivatives), or control a single degree of freedom, which
sidesteps this problem.
\item Use non-neural systems to produce motor commands.
\item Do not model a biologically plausible form of synaptic learning.
\end{enumerate}

For these reasons the approach we presented towards motor learning may be the
most capable yet, in its ability to self-configure actuators while still
respecting a large amount of biological constraints. Moreover, there is a clear
vision on how to extend this model so it can tackle more complex tasks and
controllers.

\section{Conclusion}
In this paper we have introduced the main ideas required for a class of motor
control models that maintain a large degree of biological plausibility, while
still being capable of performing non-trivial tasks. There are 3 key
characteristics that make this possible: a feedback control architecture using dual
excitatory-inhibitory representations, synaptic rules that find the direction of
sensitivity derivatives, and a critic component that configures the controller
using reinforcement learning mechanisms.
The fact that these models
have hierarchical extensions that could potentially be used to control
homeostatic variables opens the possibility of our ideas
producing highly adaptable autonomous agents. We will work towards this goal.

\section*{Acknowledgments}
The authors want to thank Prof. Kenji Doya for numerous and helpful comments to
initial versions of this manuscript.

\section*{Supplementary Material}
The source code for this paper can be obtained from:

\url{https://gitlab.com/sergio.verduzco/public_materials} \\
in the \verb+synaptic_approach+ folder.


\appendix

\section{Analogy with the Relative Gain Array Criterion.}
\label{app:rga}
When presenting equation \ref{eq:rga} in section \ref{sub:rules}
it was mentioned that this has similarities to the Relative Gain Array (RGA)
criterion. We explain that comment.

Assume a Multi-Input Multi-Output (MIMO) system where the plant is
$M$-dimensional, and the  controller has an $N$-dimensional output. Further
assume that we want to create a decentralized control system, consisting of $N$
individual feedback loops. In a control system like the one in figure
\ref{fig:neg_feedback} of the
main text, the problem we face is knowing which controller should be assigned to
control each state variable. Since control loops will be interacting with each
other, performance will be degraded, but a good loop configuration (also called
input/output selection) can largely attenuate this.

The RGA criterion \citep{bristol_new_1966} offers a measure of the interaction
between
control variables and plant outputs (or in our case, elements of the error 
vector) that, among other things, has the desirable property of scale
invariance. Consider a linearised, time-invariant control system
$ \dot{\bar{y}} = A\bar{y} + B\bar{u}$, where $\bar{u}$ is the $N$-dimensional
control vector, and $\bar{y}$ is the $M$-dimensional observed plant output. To
simplify the presentation we use a 2x2 system:
\begin{equation*}
\begin{bmatrix}
\dot{y}_1 \\
\dot{y}_2 
\end{bmatrix}
= 
\begin{bmatrix}
a_{11} & a_{12} \\
a_{21} & a_{22}
\end{bmatrix}
\begin{bmatrix}
y_1 \\
y_2 
\end{bmatrix}
+
\begin{bmatrix}
b_{11} & b_{12} \\
b_{21} & b_{22}
\end{bmatrix}
\begin{bmatrix}
u_1 \\
u_2 
\end{bmatrix}.
\end{equation*}

By assumption, the system is stable for constant $\bar{u}^*$ controls, so that
at a fixed point we have $A\bar{y}^* + B\bar{u}^* = 0$. We may thus write:
\begin{equation*}
\bar{y}^* = A^{-1}B\bar{u}^* \equiv K\bar{u} =
\begin{bmatrix}
k_{11} & k_{12} \\
k_{21} & k_{22}
\end{bmatrix}
\begin{bmatrix}
u_1 \\
u_2 
\end{bmatrix},
\end{equation*}
where $K$ is a steady-state gain matrix.
The RGA method uses $K$ to produce a matrix $\Lambda$ whose entries are defined
to be:
\begin{equation*}
\lambda_{i,j} =
\frac{(\Delta y_i/\Delta u_j)_{\Delta u_j}}{(\Delta y_i/\Delta u_j)_{\Delta y_i}}.
\end{equation*}
$\lambda_{i,j}$ is a measure of the interaction between $y_i$ and $u_j$, arising
from the ratio of two gains. The gain $(\Delta y_i/\Delta u_j)_{\Delta u_j}$
is $(\Delta y_i/\Delta u_j)$ when $\Delta u_l=0$ for $l \neq j$. In other words,
this gain is produced from the plant's outputs when $\Delta u_j$ is the only
non-zero perturbation. $(\Delta y_i/\Delta u_j)_{\Delta y_i}$
is $(\Delta y_i/\Delta u_j)$ when $\Delta y_l=0$ for $l \neq i$. For example, to
find $(\Delta y_1/\Delta u_1)_{\Delta u_1}$ we set the equation:
\begin{equation*}
\begin{bmatrix}
y_1^* + \Delta y_1 \\
y_2^* + \Delta y_2 \\
\end{bmatrix}
= 
\begin{bmatrix}
k_{11} & k_{12} \\
k_{21} & k_{22}
\end{bmatrix}
\begin{bmatrix}
& u_1^* + \Delta u_1 \\
& u_2^*
\end{bmatrix},
\end{equation*}
finding that $\Delta y_1 = k_{11} \Delta u_1$, so 
$(\Delta y_1/\Delta u_1)_{\Delta u_1} = k_{11}$.

To find $(\Delta y_1/\Delta u_1)_{\Delta y_1}$ we set
\begin{equation*}
\begin{bmatrix}
y_1^* + \Delta y_1 \\
y_2^*\\
\end{bmatrix}
= 
\begin{bmatrix}
k_{11} & k_{12} \\
k_{21} & k_{22}
\end{bmatrix}
\begin{bmatrix}
& u_1^* + \Delta u_1 \\
& u_2^* + \Delta u_2 \\
\end{bmatrix}.
\end{equation*}
Some simple algebra shows that $\Delta y_1 = \left( k_{11} -
\frac{k_{12}k_{21}}{k_{22}} \right) \Delta u_1$. Therefore
$\lambda_{1,1} = k_{11} \left( k_{11} - \frac{k_{12}k_{21}}{k_{22}}
\right)^{-1}$. It is easy to show that, in general:
$\Lambda = K \otimes (K^{-1})^T$, where $\otimes$ is the element-by-element
product. It is not difficult to prove that the rows and columns of $\Lambda$ 
add to one. Moreover, $\Lambda$ is invariant to scaling of the gain in any
controller, and permutation of the controllers only causes the same permutation
in $\Lambda$. Some stability properties of the controller can be proven when
integral action dominates, but these are not the focus of the
current exposition.

Returning to our 2x2 example, we had calculated $\lambda_{1,1} = k_{11} \left( k_{11} - 
\frac{k_{12}k_{21}}{k_{22}} \right)^{-1}$. The appearance of $k_{11}$ is simple
to interpret: it is the ratio of the reaction $\Delta y_i$ divided by the
perturbation $\Delta u_j$, as implied by the steady state gain matrix.
This ratio of reaction to perturbation could be captured in a learning rule
where $\dot{\omega}_{ij} = -\alpha \dot{e}_i(t) \dot{u}_j(t - \Delta t)$.
This would work if controllers didn't interact (e.g. columns of $K$ only had a
single non-zero element), but in general the action of one controller may
disrupt the action of the others.

To handle interaction among controllers  the RGA criterion considers the vector
$\bar{c}_i$ that is orthogonal to every row of $K$, save for the $i$-th one.
If we wanted to control $y_i$ without perturbing any other variable, then a
control output along the direction of $\bar{c}_i$ could do this.
The term $\left( k_{11} - \frac{k_{12}k_{21}}{k_{22}} \right)$ is the value of
the first entry in $\bar{c}_i$ for the 2x2 case. It would be ideal if this value
was of the same magnitude as $k_{11}$. In general, values of $\lambda_{j,k} \gg 1$
are a sign that the $k$-th controller would cause excessive interference if used
to control the $j$-th variable, whereas $\lambda_{j,k} \ll 1$ indicates that
this controller has little effect on $y_j$.

It is not obvious how to calculate $(\Delta y_i/\Delta u_j)_{\Delta y_i}$
using a biologically-plausible network. Instead, we could approximate
$\lambda_{j,k}$ by making the weight of the connection from $e_j$ to $c_k$
increase according to how much $e_j$ changes in reaction to $c_k$,
but downgrade or upgrade this increase according to how much the other
controllers are also changing $e_j$. This is the aim of the synaptic competition
introduced in equation \ref{eq:rga} of the main text.

\section{Alternative learning rules for monotonic control}
\label{app:alternative}
The rules we derive here have a
Hebbian-like form where the synaptic weight $\omega_{ij}$ for the connection 
from $e_j$ to $c_i$ has a time derivative:
\begin{equation} \label{eq:hebb_like}
        \dot{\omega}_{ij}(t) = 
            -\alpha G_j(\mathbf{e}(t)) H_i(\mathbf{c}(t)),
\end{equation}
where $\alpha$ is a learning rate, and $G_j, H_i$ are delay-differential
operators. In this appendix we present two more equations of this
type, and show a test of their performance.

For a different approach to produce a learning rule, consider using some form of
reinforcement learning in order to train the controller of figure
\ref{fig:bio_feedback}. We have to consider that the method we choose has to
act in continuous time, learn on-policy (e.g. as it performs its task), and
result in the adjustment of the $\omega_{ij}$ weights.

As a first consideration, providing rewards only when $S_D = S_P$ could result
in very slow learning, so a form of reward shaping is desirable. For this
purpose we can use $||\mathbf{e}||$ as a measure of distance to the target,
which can be a negative reward.
Training individual synapses can be addressed by a policy gradient method, with
synaptic weights being the parameter, and presynaptic rates being the state.
The weight perturbation method \cite{werfel_learning_2005} uses this logic:
a perturbation in the $\omega_{ij}$ weights causes a change in the error, which
allows to estimate the gradient of the error with respect to the weights.
As pointed out in \cite{werfel_learning_2005}, weight perturbation can be much
slower than node perturbation, which can still be relatively slow
when used to find sensitivity derivatives \cite{abdelghani_sensitivity_2008}.

To explain the node perturbation scheme (as in the REINFORCE framework
\cite{williams_simple_1992}), consider a linear system with M-dimensional
inputs $x$, and $N$ dimensional outputs $y$, related by an NxM weight matrix
$W$, so that $y = Wx$. For each input $x$ we have a desired output $d$, and we
assume that there is a teacher matrix $W^*$ such that $d = W^*x$. The error
function is $E = \frac{1}{2}|y-d|^2 = \frac{1}{2}|(W-W^*)x|^2 =
\frac{1}{2}|\Delta W x|^2$, where $\Delta W \equiv W-W^*$.

Node perturbation consists of adding noise to the outputs $y$ so we can get a
new error $E'_{NP}$, and then we change the weights following the gradient of
that error. More precisely, let $\xi$ be an N-dimensional vector drawn from a
Gaussian distribution with 0 mean and variance $\sigma^2$. Define
$E'_{NP} = \frac{1}{2}|\Delta W x + \xi|$. Weights are changed according to
$\Delta W_{NP} = -\frac{\alpha}{\sigma^2}(E'_{NP} - E)\xi x^T$.

A problem that comes with an on-policy, continuous-time implementation of this
would be to produce the same inputs twice so we can observe the error gradient 
$(E'_{NP} - E)$ using errors with and without the $\xi$ output perturbation.
The scheme we propose is to use $\dot{c}$ as a proxy for $\xi$, and
$\frac{d ||\mathbf{e}(t)||}{dt} \equiv ||\mathbf{e}(t)||'$ as a proxy for 
$(E'_{NP} - E)$, leading to a rule like:

\begin{equation*}
\dot{\omega}_{ij} = -\alpha ||\mathbf{e}(t)||' \dot{c}_i(t-\Delta t)
              e_j(t - \Delta t).
\end{equation*}

Simulations show that this rule is still not effective. In the first place, the
$e_j$ inputs are always positive, which is unlike node perturbation in the
REINFORCE framework. This
can be addressed by using the term $(e_j(t - \Delta t) - \langle \mathbf{e}
\rangle)$ instead, where $\langle \mathbf{e} \rangle = \sum_k e_k(t - \Delta t)$.
Secondly, this rule tends to produce much better results when
heterosynaptic competition is also introduced for the $\dot{c}_i$ term, in as in
the previous cases. The rule we will test in this paper is thus:

\begin{equation} \label{eq:np}
\dot{\omega}_{ij} = -\alpha ||\mathbf{e}(t)||'
                    \big( \dot{c}_i(t-\Delta t) - \langle \dot{c} \rangle \big)
                    \big( e_j(t - \Delta t) - \langle \mathbf{e} \rangle \big) .
\end{equation}

We will derive one final rule. To
understand it we must first consider that the units in population $C$ may act as
integrators of their input, a design that is justified in 
\ref{app:nonconvergence}, but can
also be understood from the following discussion.
We will write simplified equations
for the system of figure \ref{fig:neg_feedback}. Assume that the plant
$P$ is linear, with an output $\mathbf{p} = W_P \mathbf{c}$. 
Let the output of the $C$ population consist of the vector 
$\int W (\mathbf{s}_D(\xi) - \mathbf{s}_P(\xi)) d\xi$,
where $W$ is a matrix of synaptic weights, for which we want to find a learning
rule. If we assume the transmission delays and latencies of the system can
be absorbed into the response latency of the $S_P$ population with dynamics
$\tau_s \dot{\mathbf{s}}_P(t) = \mathbf{p} - \mathbf{s}_P$, the simplified
system's equations can be written as:
\begin{align}
        & \tau_s \dot{\mathbf{s}}_P(t) = W_P \int^t_0 W 
        (\mathbf{s}_D(\xi) - \mathbf{s}_P(\xi)) d\xi - 
        \mathbf{s}_P(t), \label{eq:sp} \\
        & \tau_w \dot{W}(t) = G(\mathbf{s}_D(t) - \mathbf{s}_P(t))
        H\left(\int_0^t W (\mathbf{s}_D(\xi) - 
        \mathbf{s}_P(\xi)) d\xi \right) ,
\end{align}
where $G, H$ are the matrix versions of the operators in equation
\ref{eq:hebb_like}. We would like to have a stable fixed point such that
$\mathbf{s}_D = \mathbf{s}_P$ (looking at equation \ref{eq:sp}, this fixed point
may not make sense without integration in the units of $C$). This implies there is
a time $t^*$ so that $\mathbf{e}(t^*) = \mathbf{s}_D(t^*) - \mathbf{s}_P(t^*) 
\approx \mathbf{0}$. Stability of this fixed point should imply that if there is a
perturbation $\delta \mathbf{e}$ away from the fixed point $\mathbf{e}=\mathbf{0}$,
then the control signal would move $\mathbf{e}(t)$ back towards $\mathbf{0}$,
which would be the case if $\dot{\mathbf{e}}(t+\Delta t) \approx -\delta 
\mathbf{e}$ for some small enough $\Delta t$.
A different way to state this condition is that the vector
$\mathbf{c}(t) = \int_0^{t} W(\xi) \mathbf{e}(\xi) d\xi$ is in the space
generated by all eigenvectors of $W_P$ with negative eigenvalues. Assuming
$\mathbf{e}(0) = \mathbf{0}$, it is necessary that $W(\xi) \mathbf{e}(\xi)$ is
also in this negative eigenspace for most values of $\xi \in (0,t)$.

In short, we want to modify $W(t)$ so that $W_P W(t)\mathbf{e}(t)$ 
aligns with $-\mathbf{e}(t)$. A measure of that alignment can come from the
inner product $\mathbf{e}(t) \cdot \dot{\mathbf{e}}(t+\Delta t)$, where $\Delta
t$ is roughly the time it takes the $\mathbf{e}$ signal to go through a loop in
the feedback system, causing a change $\dot{\mathbf{e}}$. In practice it is
better to use
$\mathbf{e}(t+\Delta t) \cdot \dot{\mathbf{e}}(t+\Delta t)$, which states that
we want the controller response to correct the $\mathbf{e}(t+\Delta t)$ error,
rather than the outdated $\mathbf{e}(t)$ error. Ideally we would like
to have this inner product close to its minimum value
$-||\mathbf{e}|| ||\dot{\mathbf{e}}||$. On the other hand, when the inner 
product is positive, we want to change $W$ so it produces the opposite change.

Assume that an error signal $\mathbf{e}(t)$ causes a controller response
$\dot{\mathbf{c}}(t+\Delta_0 t)$, and this activity eventually creates a change in the
error signal with rate $\dot{\mathbf{e}}(t+\Delta_2 t)$. When the inner product
$\mathbf{e}(t+\Delta_2 t) \cdot \dot{\mathbf{e}}(t+\Delta_2 t)$ is positive, we
can attempt to reduce it by subtracting the outer product 
$\dot{\mathbf{c}}(t+\Delta_0) \mathbf{e}^T(t)$ from $W$, leading to the following rule:

\begin{equation*}
        \tau_\omega \dot{\omega}_{ij}(t) = -\alpha \left[ \mathbf{e}(t) \cdot
        \dot{\mathbf{e}}(t) \right] \dot{\mathbf{c}}_i(t - \Delta_1 t)
        \mathbf{e}_j(t - \Delta_2 t).
\end{equation*}

In this equation the time shifts were made negative to avoid using future
values. $\Delta_1 t$ is the time it takes for the activity in $C$ to cause a
change in the input to $C$, and $\Delta_2 t$ is $\Delta_1 t$ plus the response
latency in $C$. The result is similar to equation \ref{eq:np} with a different
measure of the error gradient. As before, from the
activities in $\dot{\mathbf{c}}$ and $\mathbf{e}$ we can subtract the mean values
$\langle \dot{\mathbf{c}} \rangle = \sum_k \dot{c}_k $, and 
$\langle \mathbf{e} \rangle = \sum_k e_k$:
\begin{equation} \label{eq:meca_hebb}
        \tau_\omega \dot{\omega}_{ij}(t) = -\alpha \left[ \mathbf{e}(t) \cdot
        \dot{\mathbf{e}}(t) \right] 
        (\dot{\mathbf{c}}_i(t - \Delta_1 t) - \langle \dot{\mathbf{c}} \rangle)
        (\mathbf{e}_j(t - \Delta_2 t) - \langle \mathbf{e} \rangle).
\end{equation}

\begin{figure}
  \centering
  \includegraphics[width=0.9\linewidth]{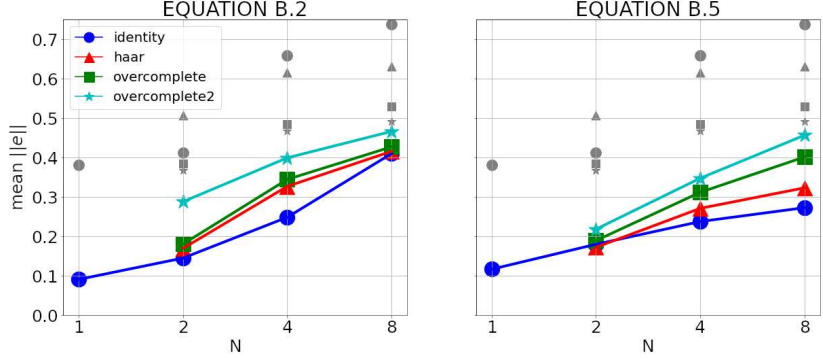}
  \caption{ Simulation results for 4 types of connectivity matrices using the
  two learning rules of this section. The format of this figure is the same as
  of figure \ref{fig:monotonic} in the main text.}
\label{fig:alternative}
\end{figure}

The rules in equations \ref{eq:np} and \ref{eq:meca_hebb} were tested with the
same linear plant as used in section \ref{sub:linear_control}, figure
\ref{fig:monotonic}. Results can be observed in figure \ref{fig:alternative}.

The worst performance is obtained with equation \ref{eq:np}, which scales
poorly to larger values of $N$. We still have to pinpoint the exact cause of
this, although we do not discard that different parameters could change the
outcome. This rule is included because it is the best adaptation we have found
of an established RL method that can be used within our framework. A general
understanding of why performance degrades will require a comprehensive
convergence analysis.

\section{Non-convergence of a simple linear model}
\label{app:nonconvergence}
In section \ref{sub:linear_control} of the main text it is stated that using the
architecture of figure \ref{fig:neg_feedback}, together with linear units and
plastic synapses as in equation \ref{eq:rga} will
lead to a network that can converge to states with non-zero error. This is shown
next for the first learning rule of section \ref{sub:rules}.

Consider the system of figure \ref{fig:neg_feedback}, with the first plant of
section \ref{sub:linear_control}. Namely,
each unit $c_j$ of the controller has a vector $\mathbf{v}_j$ associated with
it, and the output of the plant is $\mathbf{p} = \sum_k c_k \mathbf{v}_k$, where
$c_k$ is also used to denote the activity of the $k$-th unit. We define
$\mathbf{V}$ as the MxN matrix whose $N$ columns are the $\mathbf{v}_j$ vectors, 
so we may write $\mathbf{p} = \mathbf{Vc}$.

We assume that $S_P$ has an M-dimensional activity vector 
$\mathbf{s}_p = \mathbf{p}(t-d)$, where $d$ incorporates the transmission delays.
$S_{DP}$ activity is a 2M-dimensional vector with elements $s^{DP}_j =
\text{sgn(j)}(s^D_j - s^P_j)$, where $\text{sgn(j)} = 1$ for $j \leq M$, and 
$\text{sgn(j)} = -1$ for $j > M$. In vector notation this can be written
$\mathbf{s}_{DP} = \mathbf{s}_{D(2)} - \mathbf{W} \mathbf{s}_P$, where
$\mathbf{s}_{D(2)}$ is a 2M column vector with two stacked copies of
$\mathbf{s}_D$, and $\mathbf{W}$ is a 2MxM matrix consisting of the MxM identity
matrix stacked on top of its negative.

We also define $\mathbf{\Omega}$ to be the Nx2M
matrix of connections from $S_{DP}$ to $C$.

The system has the following equations

\begin{align*}
& \tau_C \dot{c}_j(t) = \left( \sum_k \omega_{jk} s^{DP}_k(t-d_1) \right) - c_j(t), \\
& \tau_\omega \dot{\omega}_{jk}(t) = (\dot{c}_j(t-d_2) - \langle \dot{\mathbf{c}}(t-d_2) \rangle )
\left( \dot{s}^{DP}_k(t - d_1) - \langle \dot{\mathbf{s}}_{DP}(t - d_1) \rangle \right)
\end{align*}
which in vector notation become:
\begin{align}
& \tau_C \dot{\mathbf{c}}(t) = \mathbf{\Omega s}_{DP}(t-d_1)  - 
\mathbf{c}(t), \label{eq:vecc} \\
& \tau_\omega\dot{\mathbf{\Omega}}(t)  = \left( \mathbf{I_N}  - 
\frac{1}{N} \mathbf{1_N}  \right) \dot{\mathbf{c}}(t-d_2)
\left[ \left( \mathbf{I_{2N}} - \frac{1}{2M} \mathbf{1_{2M}}\right) 
\dot{\mathbf{s}}_{DP}(t - d_1) \right]^T, \label{eq:vecom}
\end{align}
where $\mathbf{I_N}$ is the NxN identity matrix, $\mathbf{1_N}$ is the NxN matrix
where all entries are 1, $\mathbf{I_{2M}}$ is a Mx2M matrix of the form 
$[\mathbf{I_M} \mathbf{I_M}]$, and $\mathbf{1_{2M}}$ is an Mx2M matrix of the
form $[\mathbf{1_M}\mathbf{1_M}]$.

Equation \ref{eq:vecom} is proportional to derivatives on both sides, and will
vanish in steady state.
Also, from equation \ref{eq:vecc} it is evident that if
$\mathbf{s}_{DP}=\mathbf{0}$  in the steady state, this implies $\mathbf{c} =
\mathbf{0}$, which in turn implies $\mathbf{s}_D = \mathbf{0}$. Clearly this is
not a general solution.

The actual fixed point can be found by replacing
$\mathbf{s}_{DP}$ with $\mathbf{s}_{D(2)} - \mathbf{WVc}$ in
equation \ref{eq:vecc}:
\begin{equation*}
\tau_C \dot{\mathbf{c}}(t) = \mathbf{0} =  \mathbf{\Omega s}_{D(2)}(t-d_1) - \left(
\mathbf{\Omega WV} - \mathbf{I_N} \right) \mathbf{c}(t),
\end{equation*}
so $\mathbf{c} = 
\left(\mathbf{\Omega WV} - \mathbf{I_N} \right)^{-1}\mathbf{\Omega s}_{D(2)}$.
Whether this fixed point is attractive depends on the eigenvalues of the system
of equations \ref{eq:vecc}, \ref{eq:vecom}, where \ref{eq:vecc} is used to write
$\mathbf{\dot{c}}$ and $\dot{\mathbf{s}}_{DP}$ in terms of $\mathbf{c}$. This
analysis, however, would provide little further insight.

One final point is that equation \ref{eq:vecom} shows that homogeneous
derivatives will cause similar changes for all weights, reducing learning in
the network. It is thus necessary to avoid
synchronization, which is aided by the use of heterogeneous parameters for the
sigmoidal units, as well as heterogeneous oscillation frequencies for the
controllers (see Methods).

\section{Simulations of the pendulum with gravity}
\label{app:gravity}
Simulation of the systems in section \ref{sub:nonlinear} 
was done when the pendulum experienced gravity. All
other parameters in the system were identical, with the exception of the input
gain of the plant, which was increased (see \ref{app:parameters}).

\begin{figure}
  \centering
  \includegraphics[width=1.0\linewidth]{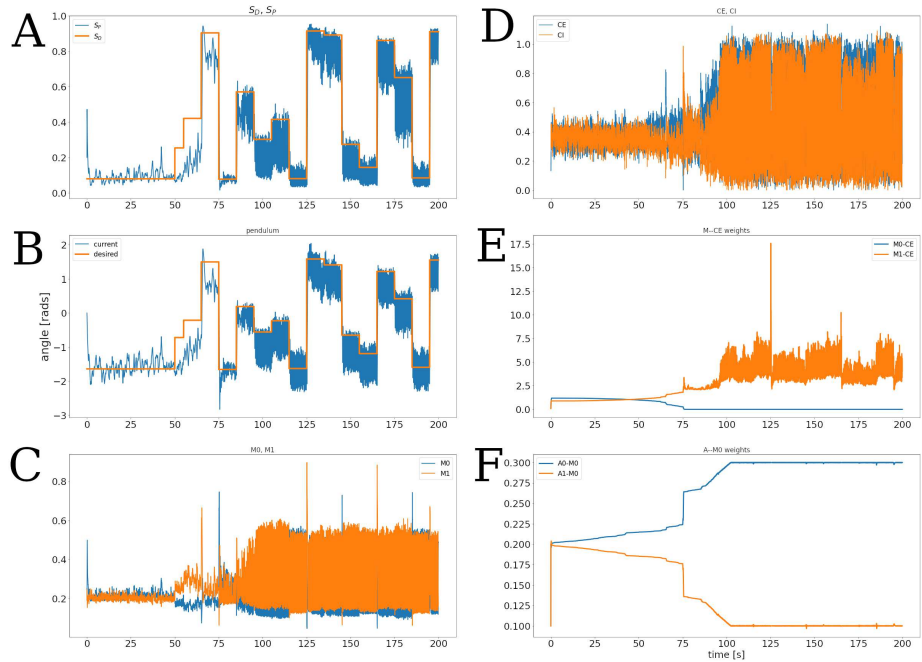}
  \caption{ First 200 seconds of a simulation where the architecture of figure
  \ref{fig:arch_pend} is used so a pendulum can track a desired angle
  (gravity is present).
  A) Activity of the $S_P$ unit, with the perceived angle, and $S_D$, with the
  desired value for $S_P$. B) Angle of the pendulum, and the desired angle.
  C) Activities of the two units in population $M$. D)
  Activities of the two units in population $C$. E) Synaptic weights for the
  connections from the two $M$ units to the $CE$ unit. F) Synaptic weights for
  the connections from the two $A$ units to one of the $M$ units.
  }
\label{fig:pendulum_2}
\end{figure}

Figure \ref{fig:pendulum_2} is the analog of figure \ref{fig:pendulum_1}, but
simulated with gravity.

The gravity force accelerates the pendulum towards the angle $-\pi/2$. Target
angles near this value require a lower gain in order to be reached, whereas
targets near $0$ and $\pi$ require a larger gain. As explained in section
\ref{sub:nonlinear}, the system is driven by error, and as the error decreases
the torque produced is not sufficient to fully reach the target. It is for this
reason that for
certain targets the system presents a larger steady-state error

The gain of the system is closely related to the slope in the sigmoidal
activation functions for the units in the feedback loop. Steep slopes will
produce high gain, but reduce the dynamic range of the system. In other words,
when the slope is steep, large or small angles will produce values very close to
1 or 0 respectively, and angle differences in these ranges will not be
perceived, leading to imprecise responses. This is a challenge that is often not
addressed by non-neural models.

A related problem with the system in section \ref{sub:nonlinear} is that for certain
initial conditions the pendulum will initially rotate close to $\pi$ radians
until the forces that restrict pendulum rotation bring it to a stop. At this
point movement of the pendulum is very limited, slowing down learning. This is
compounded by the fact that close to $\pi$ the angle is at a range where, due
to the sigmoidal activation functions, oscillation amplitudes are greatly
reduced. Both of these factors can make learning extremely slow, and leave the
pendulum ``stuck'' at $\pi$ or $-\pi$ radians.

Certain parameter regimes can be used to avoid this problem. In particular,
very fast learning rates in the connections from $M$ to $C$, 
large plant input gains, and desired values that change slowly, all contribute
to make the arm avoid getting stuck at a limit angle. This motivated some of the
parameter selection, and for this reason the first target presentation lasts 50
seconds, whereas the subsequent targets are presented for 10 seconds. It 
should be noticed that this problem is a particular
trait of using a pendulum as our test system, and is probably not relevant in a
biological setting.

When the pendulum is not constrained in its rotation this problem once again
emerges, although with a different form. When the pendulum crosses $\pi$ radians
the $s_P$ value experiences a sudden shift between 0 and 1. For certain initial
conditions, crossing $\pi$ causes the error to change its sign, making the
pendulum return to $\pi$, again changing the sign of the error. The
result is that the pendulum oscillates around $\pi$ radians indefinitely. The
same parameter regimes as before can help avoid this problem, which is once more
irrelevant for biological learning.

\section{Parameter values} \label{app:parameters}
{\it Note}: for parameters with heterogeneous values, the reported value is the one
before noise is added. All dual populations use the same parameters.
\\
For the learning equations in section \ref{sub:rules}:
\\
\begin{tabular}{|l|c|c|c|}
\hline
{\bf Parameter} & {\bf Equations} & {\bf Sections} & {\bf Value} \\
\hline
$\Delta t$ & \ref{eq:rga}, \ref{eq:mixed}, \ref{eq:np}, \ref{eq:meca_hebb} & 
    All & 140 [ms] \\
\hline
$\alpha$ & \ref{eq:rga}, \ref{eq:mixed}, \ref{eq:np}, \ref{eq:meca_hebb} &
    \ref{sub:linear_control} & .15 \\ \cline{2-4}
&   \ref{eq:mixed} & \ref{sub:nonlinear} & 2.5 \\ \cline{3-4}
&   & \ref{sub:nonmonotonic} & .5 \\ 
\hline
$\lambda$ & \ref{eq:rga}, \ref{eq:w_norm} & All & 0.05 \\ 
\cline{2-4}
& \ref{eq:mixed}, \ref{eq:w_norm} & All & 0.03 \\ 
\hline
$\tau_f$ & \ref{eq:fast_low_pass} & 
    \ref{sub:linear_control}, \ref{sub:nonlinear}, \ref{sub:nonmonotonic} &
    10 [ms] \\ \cline{3-4} 
&   & \ref{sub:linear_control} for $\dot{c_j}$;
    \ref{sub:nonlinear} for $\dot{I}_{DP}$; & 5 [ms] \\ 
&   & \ref{sub:nonmonotonic} for $\dot{I}_{DP}$ & \\
\hline
$\tau_s$ & \ref{eq:fast_low_pass} & \ref{sub:linear_control},
    \ref{sub:nonlinear}, \ref{sub:nonmonotonic}  & 50 [ms] \\ \cline{3-4}
&   & \ref{sub:linear_control} for $\dot{e_j}$  & 
200 [ms] \\ 
\hline
\end{tabular}
\\ \\

For the model in section \ref{sub:linear_control}:

\begin{tabular}{|l|c|c|c|}
\hline
{\bf Parameter} & {\bf Equation} &  {\bf Population} & {\bf Value} \\
\hline
$\tau_s$ & \ref{eq:sig} & $S_P, S_{PD}$ & 50 [ms] \\ 
\hline
$\beta$ & \ref{eq:sig}  & $S_P$ & 1 \\ \cline{3-4}
&   & $S_{PD}$ & 4 \\ 
\hline
$\eta $ & \ref{eq:sig}  & $S_P$ & 0 \\ \cline{3-4}
&   & $S_{PD}$ & 0.4 \\ 
\hline
$\tau_x$ & \ref{eq:c_2d} & $CE, CI$ & 200 [ms] \\ 
\hline
$\tau_c$ & \ref{eq:c_2d_b} & $CE, CI$ & 200 [ms] \\ 
\hline
$\tau_p$ & \ref{eq:linear_plant} & $P$ & 50 [ms] \\ 
\hline
\end{tabular}
\\ \\ \\

For the model in section \ref{sub:nonlinear}:

\begin{tabular}{|l|c|c|c|}
\hline
{\bf Parameter} & {\bf Equation} &  {\bf Population} & {\bf Value} \\
\hline
$\tau_s$ & \ref{eq:sig} & $CE, CI, S_P, S_{PD}$ & 20 [ms] \\
\cline{3-4}
&   & $M$ & 10 [ms] \\ 
\hline
$\beta$ & \ref{eq:sig}  & $CE, CI$ & 2 \\ \cline{3-4}
&   & $M$ & 2.5 \\  \cline{3-4}
&   & $S_P$ & 1.5 \\  \cline{3-4}
&   & $S_{PD}$ & 5 \\ 
\hline
$\eta $ & \ref{eq:sig}  & $CE, CI$ & 0.2 \\ \cline{3-4}
&   & $M, S_{PD}$ & 0.5 \\ \cline{3-4}
&   & $S_P$ & 0 \\
\hline
$\tau_a$ & \ref{eq:log} & $A$ & 10 [ms] \\ 
\hline
$T$ & \ref{eq:log} & $A$ & 0 \\ 
\hline
$\alpha_{IC}$ & \ref{eq:inp_corr} & --- & 0.025 \\ 
\hline
\end{tabular}
\\ \\ \\
For the model in section \ref{sub:nonmonotonic}:

\begin{tabular}{|l|c|c|c|}
\hline
{\bf Parameter} & {\bf Equation} &  {\bf Population} & {\bf Value} \\
\hline
$\tau_s$ & \ref{eq:sig} & $CE, CI, S_P, S_{PD}, V, X$ & 20 [ms] \\
\cline{3-4}
&   & $M$ & 10 [ms] \\ 
\hline
$\beta$ & \ref{eq:sig}  & $CE, CI, S_P$ & 2 \\ \cline{3-4}
&   & $M$ & 2.5 \\  \cline{3-4}
&   & $S_{PD}, X$ & 5 \\ \cline{3-4}
&   & $V$ & 1.5 \\ 
\hline
$\eta $ & \ref{eq:sig}  & $CE, CI$ & 0.2 \\ \cline{3-4}
&   & $M, S_{PD}$ & 0.5 \\ \cline{3-4}
&   & $S_P$ & 0 \\ \cline{3-4}
&   & $V, X$ & 0 \\ 
\hline
$\tau_V$ & \ref{eq:v_dynamics} & $V$ & 20 [ms] \\
\hline
$\tau_X$ & \ref{eq:x_dynamics} & $X$ & 20 [ms] \\ 
\hline
$\alpha_V$ & \ref{eq:td} & --- & 0.005 \\ 
\hline
$\Delta t_v$ & \ref{eq:td} & --- & 3 [s] \\
\hline
$\gamma$ & \ref{eq:td} & --- & 0.6 \\
\hline
$\alpha_X$ & \ref{eq:rm_hebbian1} & --- & 0.15 \\ 
\hline
$\eta_X$ & \ref{eq:rm_hebbian2} & --- & 0.2 \\ 
\hline
$\tau_a$ & \ref{eq:log} & $A$ & 10 [ms] \\ 
\hline
$T$ & \ref{eq:log} & $A$ & 0 \\ 
\hline
$\alpha_{IC}$ & \ref{eq:inp_corr} & --- & 0.025 \\ 
\hline
$\tau_P$ & \ref{eq:sp_star} & $S_P^*$ & 10 [ms] \\ 
\hline
$b$ & \ref{eq:l} & $V, X$ & 1.59 \\ 
\hline
$\eta_1$ & \ref{eq:add_terms} & $V, X$ & 0.1 \\ 
\hline
$\eta_2$ & \ref{eq:add_terms} & $V, X$ & 0.005 \\ 
\hline
\end{tabular}
\\ \\ \\
For the model in section \ref{sub:transition}:

\begin{tabular}{|l|c|c|c|}
\hline
{\bf Parameter} & {\bf Equation} &  {\bf Population} & {\bf Value} \\
\hline
$\tau_X$ & \ref{eq:tanh} & $X$ & 20 [ms] \\ 
\hline
$\beta$ & \ref{eq:tanh} & $X$ & 5 \\ 
\hline
$\alpha_X$ & \ref{eq:rm_hebbian1} & --- & 0.4 \\ 
\hline
$\eta_X$ & \ref{eq:rm_hebbian2} & --- & 0.1 \\ 
\hline
$b$ & \ref{eq:s12} & $X$ & 5 \\

\hline
\end{tabular}
\\ \\ \\
For the models in sections \ref{sub:nonlinear}, \ref{sub:nonmonotonic}, and
\ref{sub:transition}, the plant was a homogeneous pendulum of with mass 1 [kg],
and length 0.5 [m]. The viscous friction coefficient was 1 [$kg\cdot m^2/s$],
except for section \ref{sub:transition}, where the value 0.2 [$kg\cdot  m^2/s$]
was used.  When gravity is present, its value is 9.81 [$m/s^2$]. 

For the model in section \ref{sub:nonlinear}, the gain was 4, meaning that an
input of magnitude one would produce a torque of 4 [$kg\cdot m^2/ s^2$]. The
equivalent simulation in \ref{app:gravity} used a gain of 7. Section
\ref{sub:nonmonotonic} used a gain of 2, and section \ref{sub:transition} used a
gain of 1.5 .

\end{document}